\documentclass[12pt,cits,hyper]{JINST}
\pdfoutput=1

\usepackage{booktabs}
\usepackage{setspace}
\usepackage{amsmath}
\usepackage{multirow}

\newcommand{\mum}{\ensuremath{\,\mu\mathrm{m}}}

\newcommand{\degr}{\ensuremath{^{\circ}}}

\newcommand{\minitab}[2][l]{\begin{tabular}{#1}#2\end{tabular}}

\title{Argon Purification Studies and a Novel Liquid Argon Re-circulation System}
\author{K.~Mavrokoridis$^a$\thanks{Corresponding author}, R.~G.~Calland$^a$, J.~Coleman$^a$, P.~K.~Lightfoot$^b$, N.~McCauley$^a$, K.~J.~McCormick$^a$, C.~Touramanis$^a$\\
\llap{$^a$}University of Liverpool,
Department of Physics, Oliver Lodge Bld, Oxford Street, Liverpool, L69~7ZE, UK\\
\llap{$^b$}University of York,
Department of Physics, Helsington, York, YO10~5DD,  UK

E-mail: \email{k.mavrokoridis@liverpool.ac.uk}}

\abstract{
Future giant liquid argon~(LAr) time projection chambers (TPCs)
require a purity of better than 0.1 parts per billion~(ppb) to allow the ionised electrons
to drift without significant capture by any electronegative impurities. 
We present a comprehensive study of the effects of electronegative impurity on gaseous and liquid argon scintillation
light, an analysis of the efficacy of various purification chemicals, as well as the 
Liverpool LAr setup, which utilises
a novel re-circulation purification system.
Of the impurities tested - Air, O$_2$, H$_2$O, N$_2$ and CO$_2$ in the range of between 
0.01~ppm to 1000~ppm - H$_2$O was found to have the most profound effect on
gaseous argon scintillation light, and N$_2$ was found to have the least.
Additionally, a correlation between the slow component decay time and the total energy deposited with 
0.01~ppm - 100~ppm O$_2$ contamination levels in liquid argon has been established.
The superiority of molecular sieves over anhydrous complexes at absorbing Ar gas, N$_{2}$ gas and H$_{2}$O vapour
has been quantified using BET isotherm analysis. The efficiency of Cu and P$_{2}$O$_{5}$ at 
removing O$_2$ and H$_2$O impurities from 1~bar N6 argon gas at both room temperature and -130~$^{\circ}$C was investigated and found to be high.   
A novel, highly scalable LAr re-circulation system has been developed. The complete system,
consisting of a motorised bellows pump operating in liquid and a purification cartridge, were
designed and built in-house. The system was operated successfully over many days and achieved 
a re-circulation rate of 27~litres/hour and high purity.}


\begin{document}

\section{Introduction}
Purity of liquid argon (LAr) is an essential requirement for the operation of 
any LAr time projection chamber (TPC) detector.
The ionisation electrons produced in a LAr
detector, such as ArDM~\cite{Rubbia:2006, Laffranchi:2007p1321, Regenfus:2010p3319}, a 1-ton two-phase scintillation/ionisation Dark
Matter detector,
must be able to drift over a distance longer than a metre, 
without substantial capture by electronegative impurities.  
The distances over which electrons will be required to be drifted will greatly increase
as LAr TPCs expand in size, for example the proposed
GLACIER (Giant Liquid Argon Charge Imaging ExpeRiment) neutrino TPC which will have
drift distances of up to 20 metres~\cite{Rubbia:2004p3272, Rubbia:2009p3284}.
In the ICARUS LAr TPC, a maximum drift distance of 1.5~m requires a level of 
$\lesssim$~0.1 parts per billion~(ppb) oxygen equivalent~\cite{Amerio2004329},
and in recent operations actual levels of $\lesssim$~0.05~ppb were routinely achieved~\cite{ICARUS2011}.

\begin{figure}[t]
\begin{center}
\begin{tabular}{c}
	\includegraphics[width=.65\textwidth]{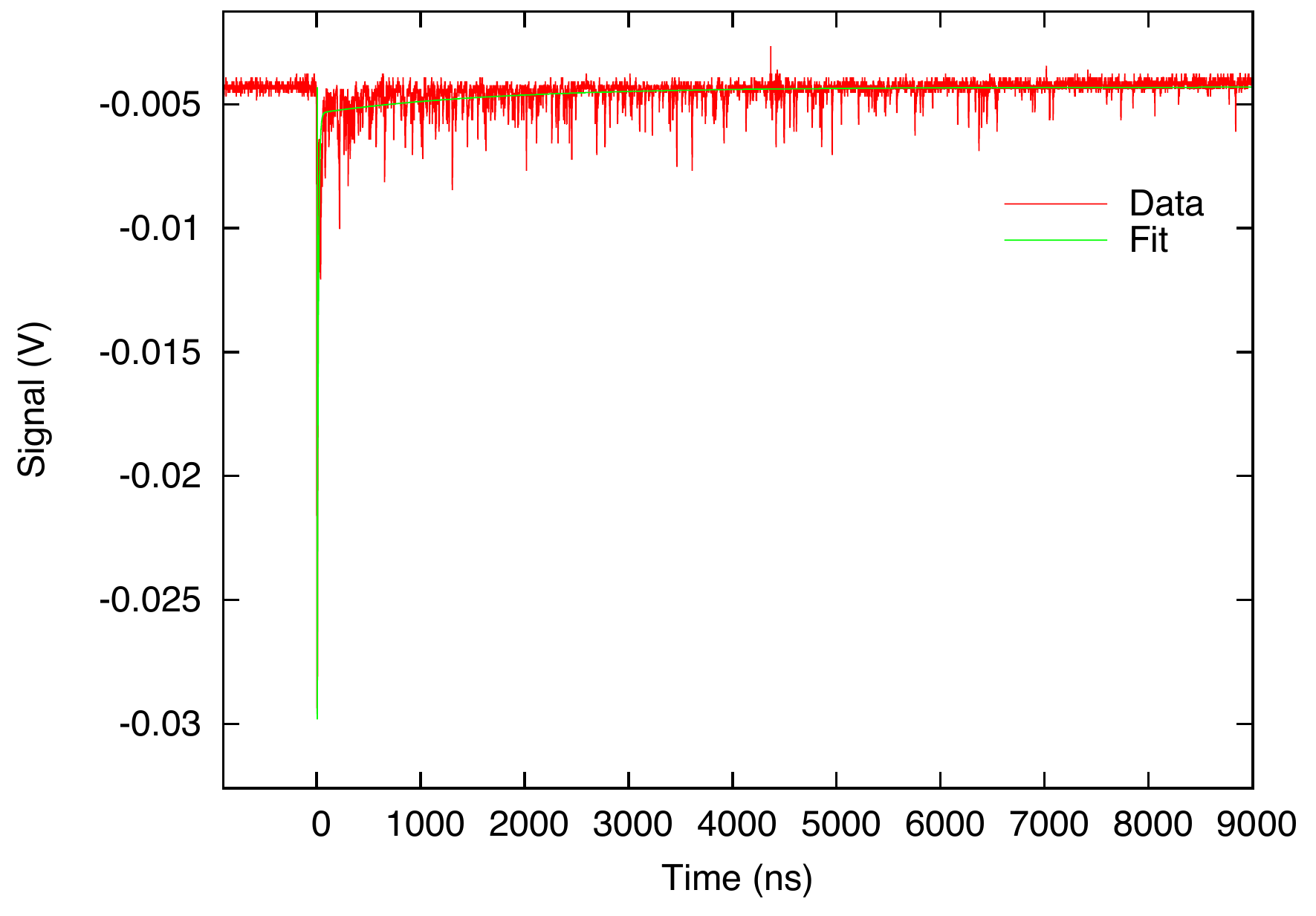}
\end{tabular}
\end{center}

\caption{A typical argon gas scintillation pulse.}

\label{pulse_argon}
\end{figure}

Argon scintillation emission is centred at approximately 128~nm
and is characterised by two distinct decay times: a slow component, $\tau_{2}$ (triplet eximer),
and a fast component, $\tau_{1}$ (single eximer)~\cite{Keto:1974p593,Kubota:1978p975}. 
To match the high quantum efficiency range of a photomultipler tube~(PMT), the argon scintillation
light can be shifted to visible using tetraphenyl-butadiene (TPB)~\cite{WLSkostas}.
A typical argon scintillation pulse at room temperature (RT) is shown in Figure~\ref{pulse_argon}.
The decay time of the slow component, $\tau_{2}$, increases with the increase of argon purity
and therefore  can be used as a relative measure of purity.  
The purest argon gas has been reported to have a $\tau_{2}$ of $ 3200\pm300$ ns \cite{Keto:1974p593},
whereas for pure liquid $\tau_{2}$ values of between 
1100~ns and 1600~ns have been reported~\cite{Morikawa1989, Hitachi1983}.
This variation has been attributed partially to different fitting methods~\cite{WarpN2contam}.

In addition to impeding the drifting of electrons, impurities within liquid argon can also absorb 
emitted UV photons or quench argon excimers, leading to a loss of light collection.
The influence of various levels of Air, O$_2$, H$_2$O, N$_2$, CO$_2$ impurity concentrations on light
collection in argon gas is examined in Section~\ref{impurities_effect}, whereas the effect of increasing O$_2$ 
impurities on LAr scintillation light, using a different experimental setup, is examined in Section~\ref{contamLAr}.
Previous studies of the effect of N$_2$ and O$_2$ contamination in LAr have been 
reported in~\cite{WarpN2contam, WarpO2Contam2010}. 

The use of materials such as molecular sieves, copper and
Oxisorb{\footnotesize$^{\textregistered}$} reagent
for gaseous argon (GAr) and LAr purification have been reported in~\cite{Maksimovich1968, Doe1987, Bressi:1990, Gennini}.
The purification mechanisms of molecular sieves are described and 
their efficiency is examined by adsorption measurements using BET~(Brunauer, Emmett, Teller)
isotherms~\cite{Brunauer:1938p2076},
in Sections \ref{mol_sieves} and ~\ref{adsorption} respectively. 
The efficiency of copper~(Cu) and
phosphorous pentoxide~(P$_{2}$O$_{5}$) at removing 
oxygen and water respectively is also assessed in Section~\ref{CuPO5}.

Re-circulation of LAr through purification chemicals, such as those described in this paper,
greatly enhances the rate and
efficiency of purification. 
Purification of LAr has traditionally been performed by re-circulation in the gas phase followed by re-condensation.
In Section~\ref{LiverpoolSetup} we present a novel re-circulation system which functions entirely in the liquid phase, and it crucially 
has the potential to be highly scalable, for use in kiloton detectors.

%%=============Argon Pulse figure===================
%%---------------------------
%\clearpage

\section{Effects of Impurities on Argon Gas Scintillation Light} \label{impurities_effect}
\subsection{Experimental Procedure}
\begin{figure}[t]
\begin{center}
\begin{tabular}{c}
	\includegraphics[width=.49\textwidth]{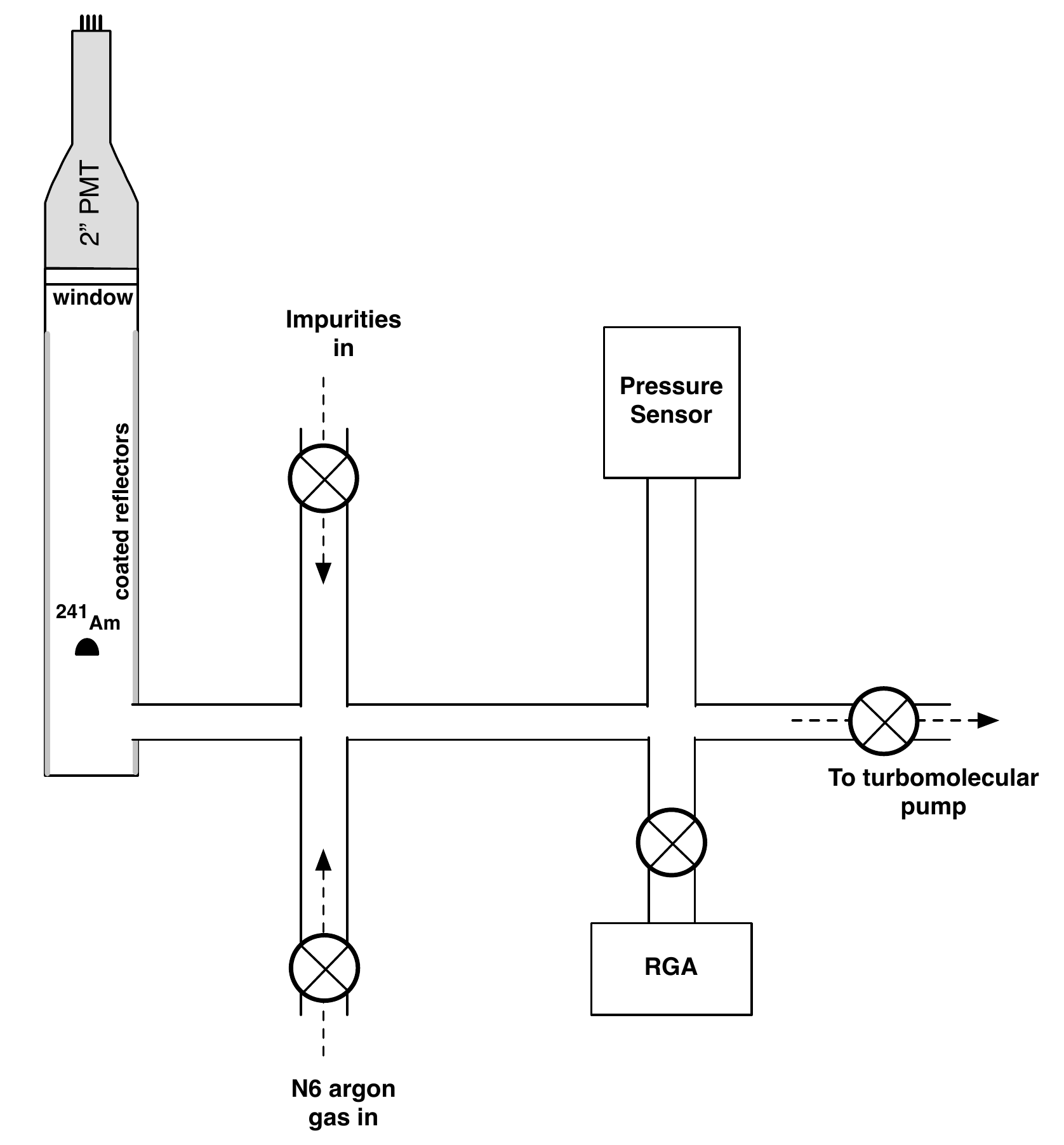}
\end{tabular}
\end{center}

\caption[Schematic of the experimental apparatus used to measure the effect of electronegative impurities on
gaseous argon scintillation light]
{ A schematic illustration of the experimental apparatus used to measure the effect of electronegative impurities on gaseous argon scintillation light.}

\label{ipmpur_aparatus_schem}
\end{figure}

As a prelude to liquid argon purity measurements, the effect of introducing the contaminants air,
oxygen, carbon dioxide, nitrogen, and water vapour into a highly purified 1 bar GAr environment (no gas flow) was measured. 
A schematic illustration of the experimental apparatus is shown
in Figure~\ref{ipmpur_aparatus_schem}.
N6 argon (i.e. 1~ppm impurity level) was flowed at 2~lpm through a SAES getter which was nominally able to purify rare gases containing oxygen, water and carbon dioxide to 1~ppb levels. Each contaminant was introduced individually via a separate valve, and a residual gas analyser~(RGA) connected through a bleed valve was used to measure the partial pressures of each element. 
The effect of the introduced impurities on GAr scintillation light was measured using a  2-inch ETL 9831KB PMT fixed
externally to a viewing window at the top of a 600~mm long DN40 tube 
featuring 1~mg/cm$^{2}$ TPB coated 3M{\small\texttrademark}-foil walls and an alpha source at the base.
The PMT signal was digitised at a sampling rate of 1 GS/s, using an Acqiris DP1400 digitiser.
 As the concentration of each impurity was varied, the PMT data was recorded and analysed for the fast component area, fast component time constant, total area, and slow component time constant.
%\clearpage

%%------------------------------

\clearpage

\subsection{Results \& Conclusions}

\begin{figure}[b]
\begin{center}
\begin{tabular}{c}
	\includegraphics[width=.72\textwidth]{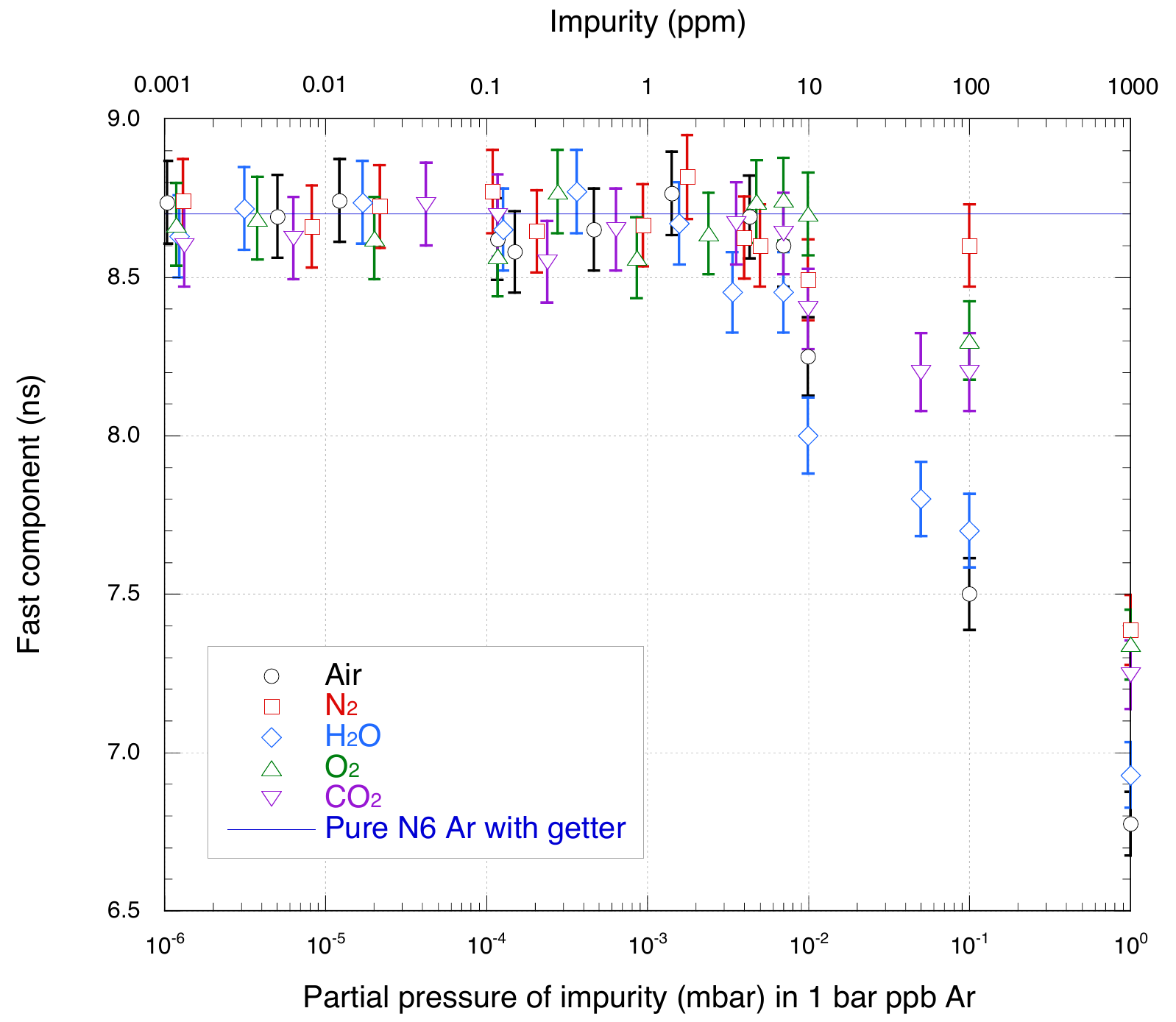}
\end{tabular}
\end{center}

\caption{Argon gas purity: Fast component (singlet) decay time variation with contaminant
partial pressure.}

\label{fastcomp_impur}
\end{figure}

\begin{figure}[t]
\begin{center}
\begin{tabular}{c}
	\includegraphics[width=.72\textwidth]{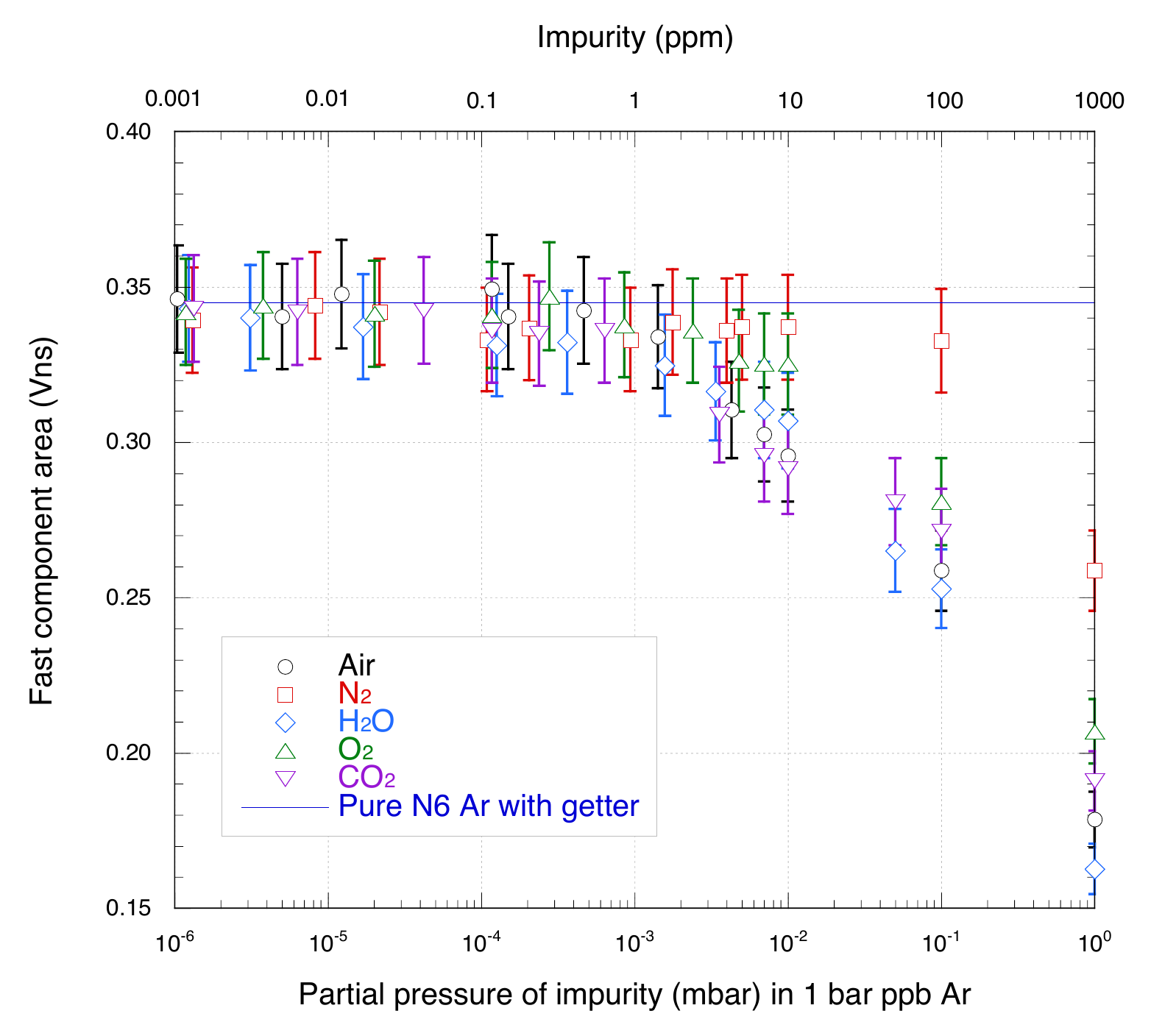}
\end{tabular}
\end{center}

\caption{ Argon gas purity: Fast component (singlet) area variation with contaminant
partial pressure.}

\label{fastarea_impur}
\end{figure}

Figures~\ref{fastcomp_impur} and \ref{fastarea_impur} show that the fast component
is only significantly affected by impurities at partial pressures greater than
10$^{-2}$~mb. As 1~bar commercial N6 argon gas has an impurity partial pressure 
of 10$^{-3}$~mb, the effect of impurities on the fast component is not a concern.

\begin{figure}[t]
\begin{center}
\begin{tabular}{c}
	\includegraphics[width=.72\textwidth]{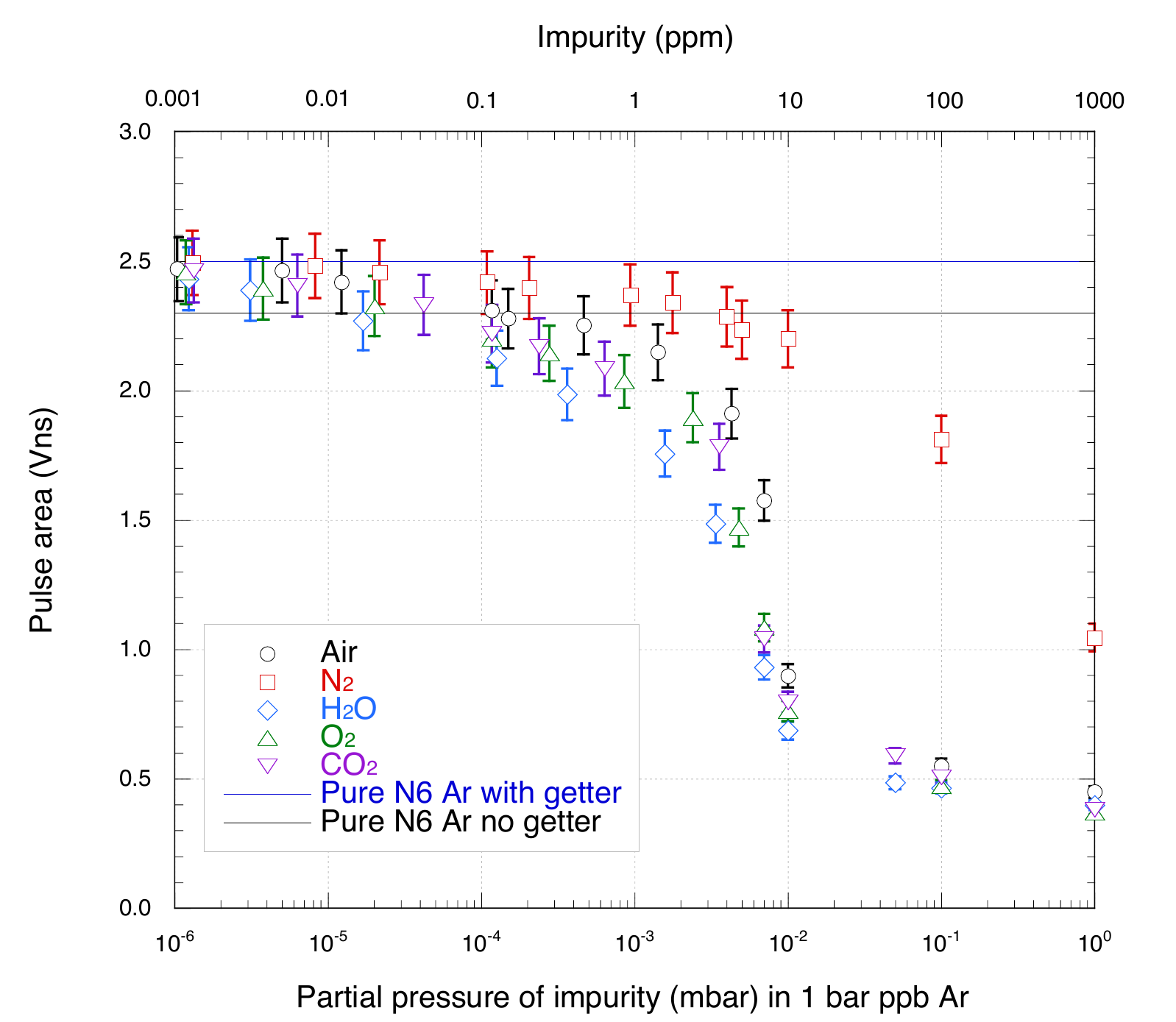}
\end{tabular}
\end{center}

\caption{ Argon gas purity:  Total pulse area variation with contaminant
partial pressure.}

\label{pulsearea_impur}
\end{figure}

\begin{figure}[t]
\begin{center}
\begin{tabular}{c}
	\includegraphics[width=.72\textwidth]{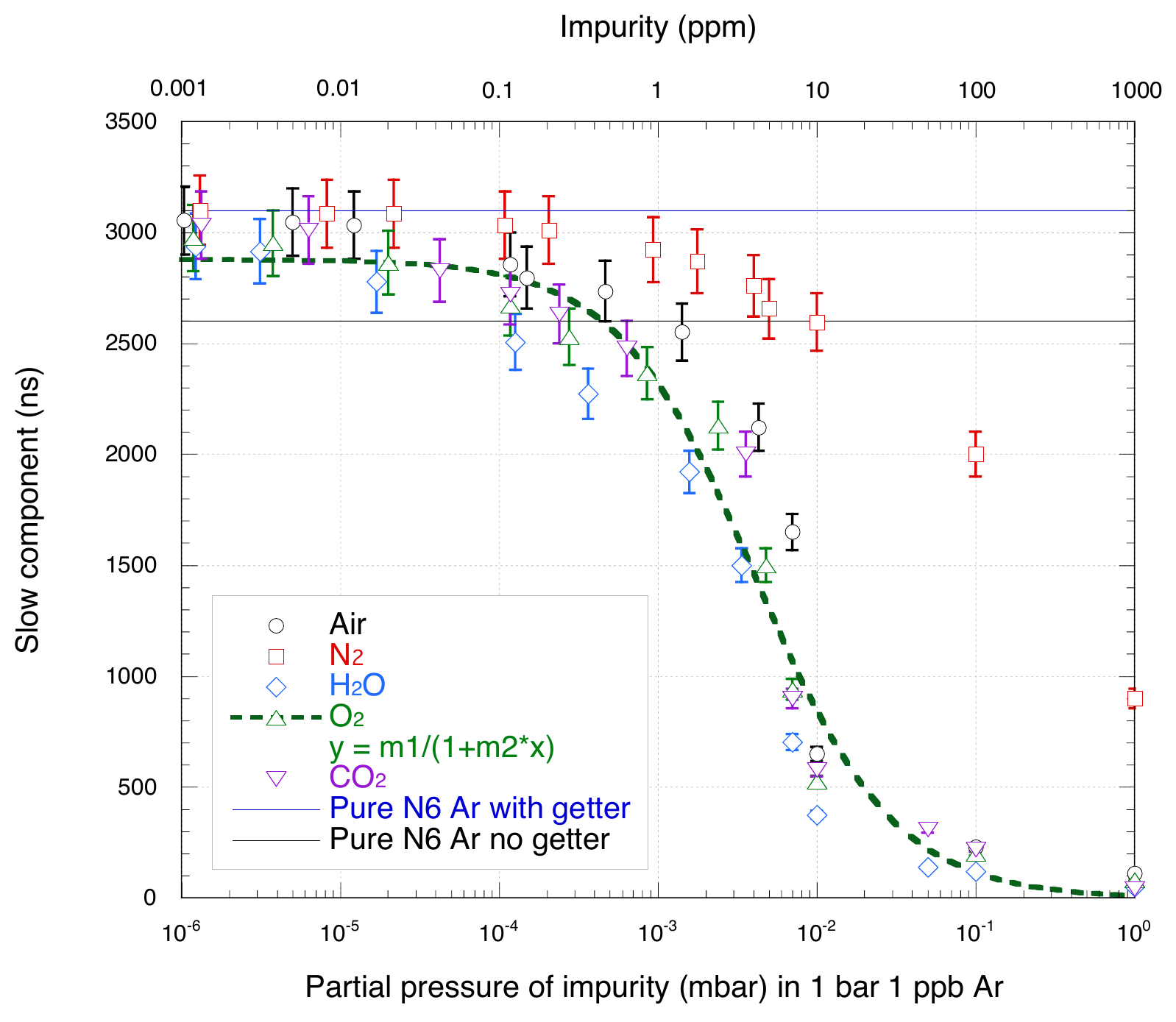}
\end{tabular}
\end{center}

\caption{ Argon gas purity:  Slow component (triplet) decay time variation with contaminant
partial pressure. The O$_2$ data were fitted with a Birk's law type function, 
where m1~=~2878~ns$\pm$80~ns 
and m2~=~0.24$\pm$0.03.}

\label{tau2impur}
\end{figure}

However, the slow component is much more sensitive to impurities, as can be seen
in Figure~\ref{tau2impur} which shows that impurity partial pressures greater
 than 10$^{-4}$~mb result in a significant decrease 
in the slow component decay time and a
reduction in the total pulse area as shown in Figure~\ref{pulsearea_impur}.

Water was found to be very marginally the most significant impurity, followed by carbon
dioxide and oxygen.
Nitrogen was found to be the most benign impurity within argon gas, although it is
most likely a VUV absorber rather than a quencher of the triplet state.

The quenching effect of 
impurities on the slow component of the argon gas scintillation light can approximately be described by
a Birks' law type function:

\begin{equation}
\tau _2 [ns] = \frac {\tau ^\prime}{1 + k \tau _2 [ppm]}
\end{equation}
where, $\tau$$^\prime$ is the asymptotic value of $\tau _2$  for zero impurities and k is a constant.
The fitting of the O$_2$  data~(Figure~\ref{tau2impur}), yields $\tau$$^\prime$~= 2878~ns$\pm$80~ns and
k =~ 0.24$\pm$0.03.  

%\clearpage
%%------

\clearpage

\clearpage

%%------------
%\clearpage

\section{Molecular Sieves} \label{mol_sieves}

Molecular sieves  contain small pores. 
Molecules small enough to pass through the pores are trapped while larger molecules are not.
This property is called size or steric exclusion. 
Activated carbon, alumina (aluminium oxide), silica gel and zeolites are examples of molecular 
sieves~\cite{Bekkum2001}. 
Whilst all, especially activated carbon, are relatively inexpensive and suited to
capturing large organic molecules,
the latest molecular sieves such as zeolites are made of complex crystallographic structures of Na, K, Ca, Al, Si
which do not collapse when dehydrated~\cite{scott_ebook}. 
By selecting the optimal blend, molecular sieves can be graded to
capture molecules with sizes between 3~\AA\ and 10~\AA. 
In addition to steric exclusion, zeolites have an extra property called thermodynamic selectivity.
This is a measure of  preferential absorption of certain components over others when all
components are able to enter the pores. A broad majority of gas separations by zeolites occur
because of this property. Thermodynamic selectivity is achievable because the adsorption of 
a particular gas is favourable over another on the accessible cationic sites within the crystal~\cite{zeolite}.
The adsorption capacity and selectivity depends on several parameters such as the size, shape, and structure of 
the zeolite cavity, cationic charge density, concentrations of cations, polarizability, and the dipole and quadrupole
moment of the guest adsorbate molecules.
In general, molecules with a stronger quadrupole moment and higher polarizability will 
be adsorbed more strongly. For example the larger quadropole moment of N$_{2}$ 
relative to O$_{2}$ was identified as being responsible for the N$_{2}$ selectivity of many
zeolites~\cite{barrer1959}.
Molecular sieves can be used to purify argon based on the size/steric exclusion and  thermodynamic
selectivity properties.

%%%-----------------------
\section{Adsorption of N$_{2}$, Ar and H$_{2}$O}
\label{adsorption}
Measurement of the enthalpy of adsorption, total capacity, and effective surface area of interaction, for argon gas, nitrogen gas and water vapour on a range of molecular sieves and anhydrous complexes is described in this
section.
Adsorption is the process by which molecules or atoms become attached to a surface. This is particularly significant when a liquid or a gas comes into contact with a porous solid such as charcoal or aluminium oxide. The forces that bind an adsorbed molecule (adsorbate) to the surface may be chemical or physical, giving rise to chemisorption
and physisorption respectively. 
The BET isotherm~\cite{Brunauer:1938p2076} describes adsorption and provides a route to measuring both the average enthalpy of adsorption and the effective surface area of the adsorber (e.g. activated charcoal). 
The BET equation is valid for multilayers of adsorbate assuming: the absorbate is an ideal gas; 
the energy of adsorption (energy released as the adsorbate bonds to the surface) is uniform
over the entire surface; no adsorbate-adsorbate interaction occurs; 
the adsorbed molecules are immobile.

The BET equation is given by:

\begin{equation} \label{eq:BET}
\displaystyle\frac{\mathrm{x}}{\mathrm {V(1-x)}} =
\frac{\mathrm{1}}{\mathrm {V_{m}C}} + \frac{\mathrm{(C-1)x}}{\mathrm {V_{m}C}}
\end{equation}
where, x is the relative pressure (P/p$_{o}$) at which a volume of gas V, measured at room
temperature and pressure, is adsorbed. P is the pressure of the gas directly above the adsorber once equilibrium has been reached, and p$_{o}$ is the saturation vapour pressure at the temperature of the reaction chamber containing the adsorber. V$_{m}$ is the volume of adsorbed gas required for the formation of a monolayer on the adsorber at room temperature and pressure.
C is an equilibrium rate constant given by:

\begin{equation} \label{eq:C}
C =\displaystyle\frac{\mathrm{a_{1}b_{2}}}{\mathrm{a_{2}b_{1}}} e^{\frac{\mathrm{E_{1}-E_{L}}}{\mathrm{RT}}}
\end{equation}
where, E$_{1}$  is the average enthalpy of adsorption in the first layer and
E$_{L}$ is the enthalpy of liquefaction of the adsorbate. a$_{1}$, a$_{2}$, b$_{1}$ and b$_{2}$ are constants
related to the formation and evaporation of the first and higher layers of adsorbed molecules where
 a$_{1}$b$_{2}$/a$_{2}$b$_{1}$ $\approx$ 1.
The BET equation maintains a linear relationship only in the range of 0.05 $<$ x $<$ 0.35.
A plot of the left-hand
term of the BET equation against the relative pressure
x, allows evaluation of the
constants C and  V$_{m}$ from the slope and intercept.
 
The area of each adsorbed molecule can be calculated using the following equation:
 \begin{equation} \label{eq:Mol_area} 
 \mathrm{Molecular} \: \mathrm{area} =(4) (0.866) \left[\frac{\mathrm{M}}{\mathrm{4(2)^{\frac{1}{2}}Nd}}\right]^{\frac{2}{3}}
 \end{equation}
 where, M is the molecular weight of the adsorbate in kg, N is Avogadro's number and d the density 
of the liquefied adsorbate in kg/m$^{3}$.

Since V$_{m}$ is the volume of gas at room temperature and pressure required to form a monolayer, then using the ideal gas law, the corresponding number of moles required to form a monolayer can be found, and from this the mass. Since the molecular area above represents the area taken up by one molecule, the total surface area can be calculated.

\subsection{Experimental Method}

\begin{figure}[t]
\begin{center}
\begin{tabular}{c}
	\includegraphics[width=.74\textwidth]{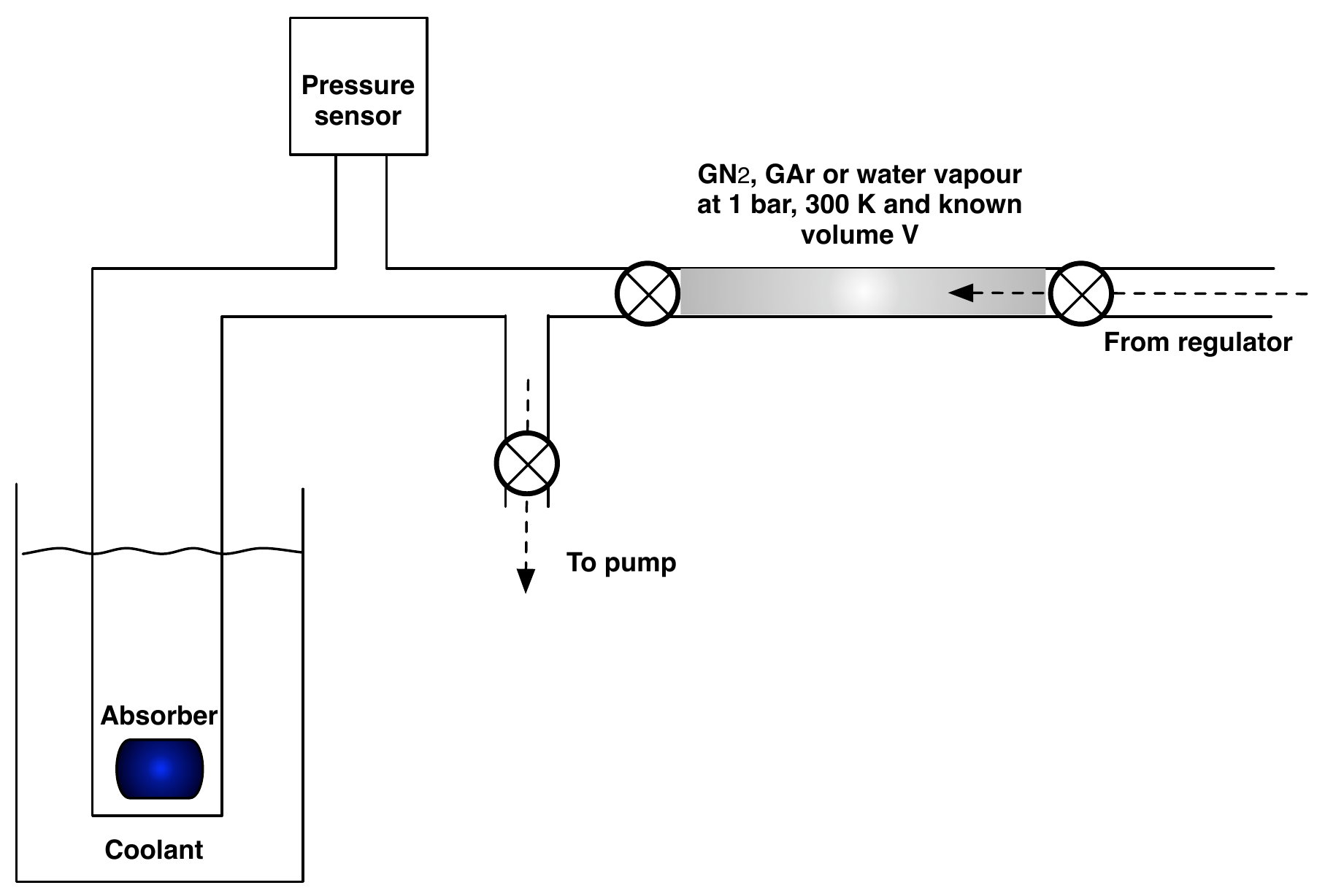}
\end{tabular}
\end{center}

\caption[Schematic of the experimental apparatus used for
adsorption measurements]
{ A schematic illustration of the experimental apparatus used to measure
the enthalpy of adsorption for argon gas, nitrogen gas and water vapour
on molecular sieves and anhydrous complexes.}
\label{BETapparatusschem}
\end{figure}

A schematic of the experimental apparatus utilised in this series of measurements 
is shown in Figure~\ref{BETapparatusschem}.
A cylinder containing high pressure adsorbate gas was connected through two valves to a chamber
containing an adsorber and a pressure sensor. 
All three valves shown in Figure~\ref{BETapparatusschem} were opened and the adsorber was vacuum pumped 
for 4 hours and heated to 600~K. 
Once the apparatus had cooled down, all valves were shut and a LN$_{2}$ dewar was placed around the adsorber chamber
(for the water vapour tests the LN$_{2}$ dewar was replaced by a cold water dewar). 
The saturation vapour pressure~(SVP) of N$_{2}$ and Ar at  LN$_{2}$ temperature
are approximately 1000~mbar and 800~mbar respectively.
The right hand valve was then opened and the test gas allowed to fill the volume marked in grey at 1 bar. The right hand valve was then closed and the middle valve opened, allowing gas to pass into the adsorber chamber. Over time (5 to 10 minutes) the pressure drops until equilibrium is reached. The gas of known volume V and the equilibrium pressure~P were then recorded. 
This procedure was repeated until the equilibrium pressure approached 35\% of the saturated vapour pressure, and the cumulative volume added was recorded against P/p$_{o}$. The adsorber was then maintained in 1 bar of gas until total saturation occurred, thus providing a measure of the total capacity.
The entire experiment was performed separately for 
nitrogen gas, argon gas and water vapour, on a range of molecular sieves and anhydrous
complexes.

%\clearpage

%\clearpage
\subsection{Adsorption Results \& Conclusions }
\begin{figure}[t]
\begin{center}
\begin{tabular}{c}
	\includegraphics[width=.8\textwidth]{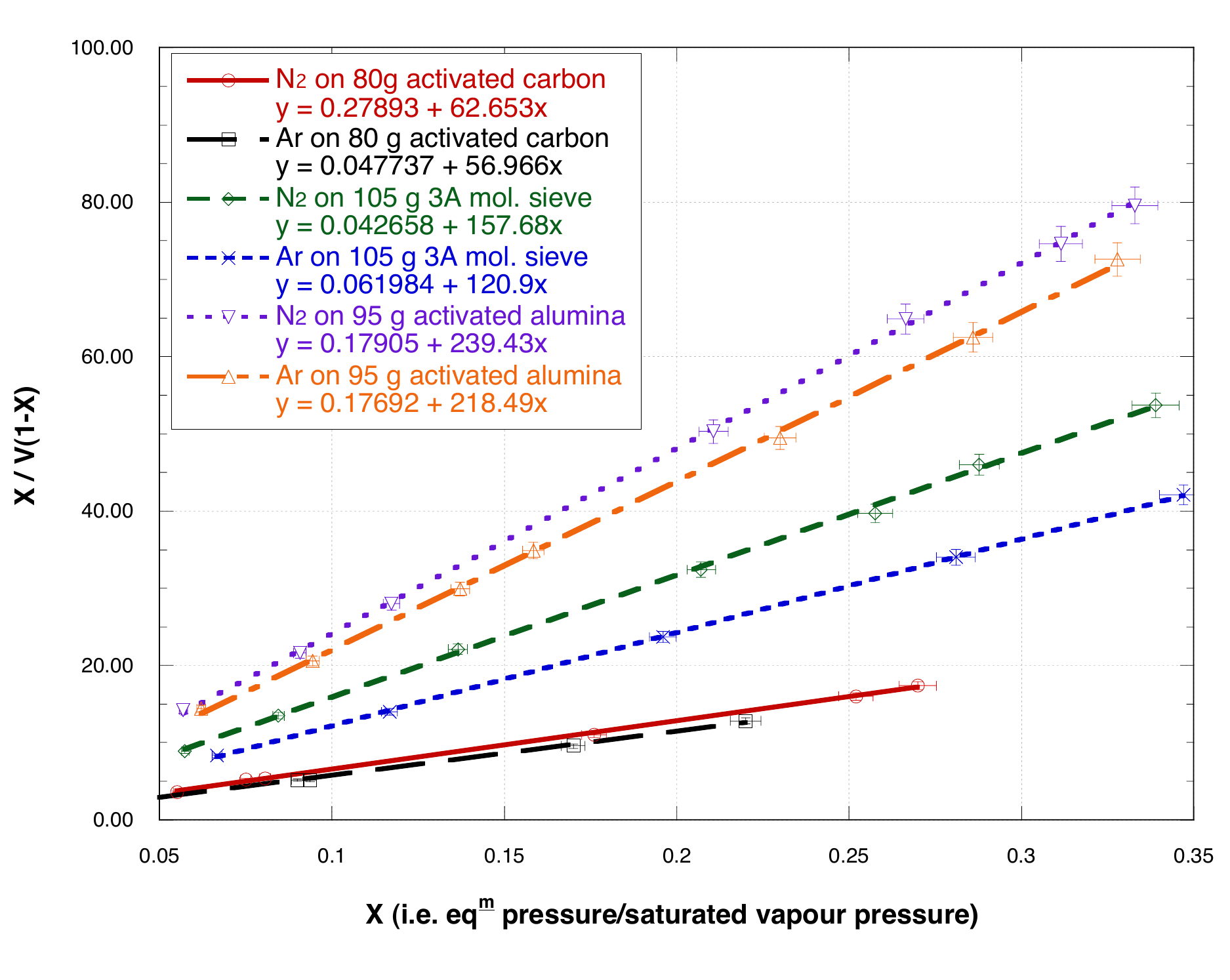}
\end{tabular}
\end{center}

\caption{Adsorption of nitrogen and argon gas on a range of adsorbers and molecular sieves.}

\label{N2_ArBET1}
\end{figure}

\begin{figure}[t]
\begin{center}
\begin{tabular}{c}
	\includegraphics[width=.8\textwidth]{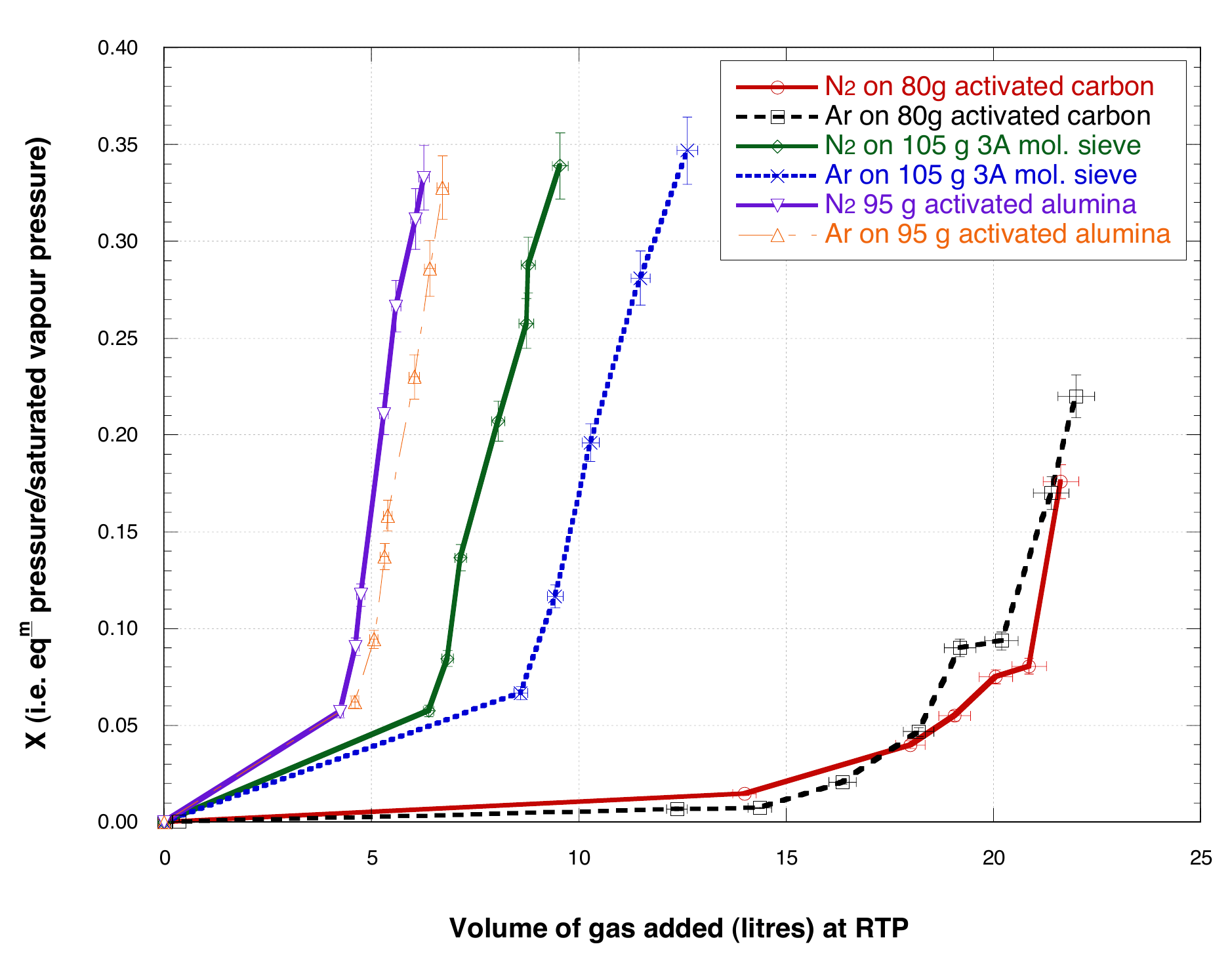}
\end{tabular}
\end{center}

\caption[Adsorption of nitrogen and argon gas on a range of adsorbers and molecular sieves
as a function of added gas]
{Adsorption of nitrogen and argon gas on a range of adsorbers and molecular sieves
as a function of added gas. Lines are present to illustrate trends and do not indicate
a fit to the data.}

\label{N2_ArBET1VOL}
\end{figure}

\begin{figure}[t]
\begin{center}
\begin{tabular}{c}
	\includegraphics[width=.8\textwidth]{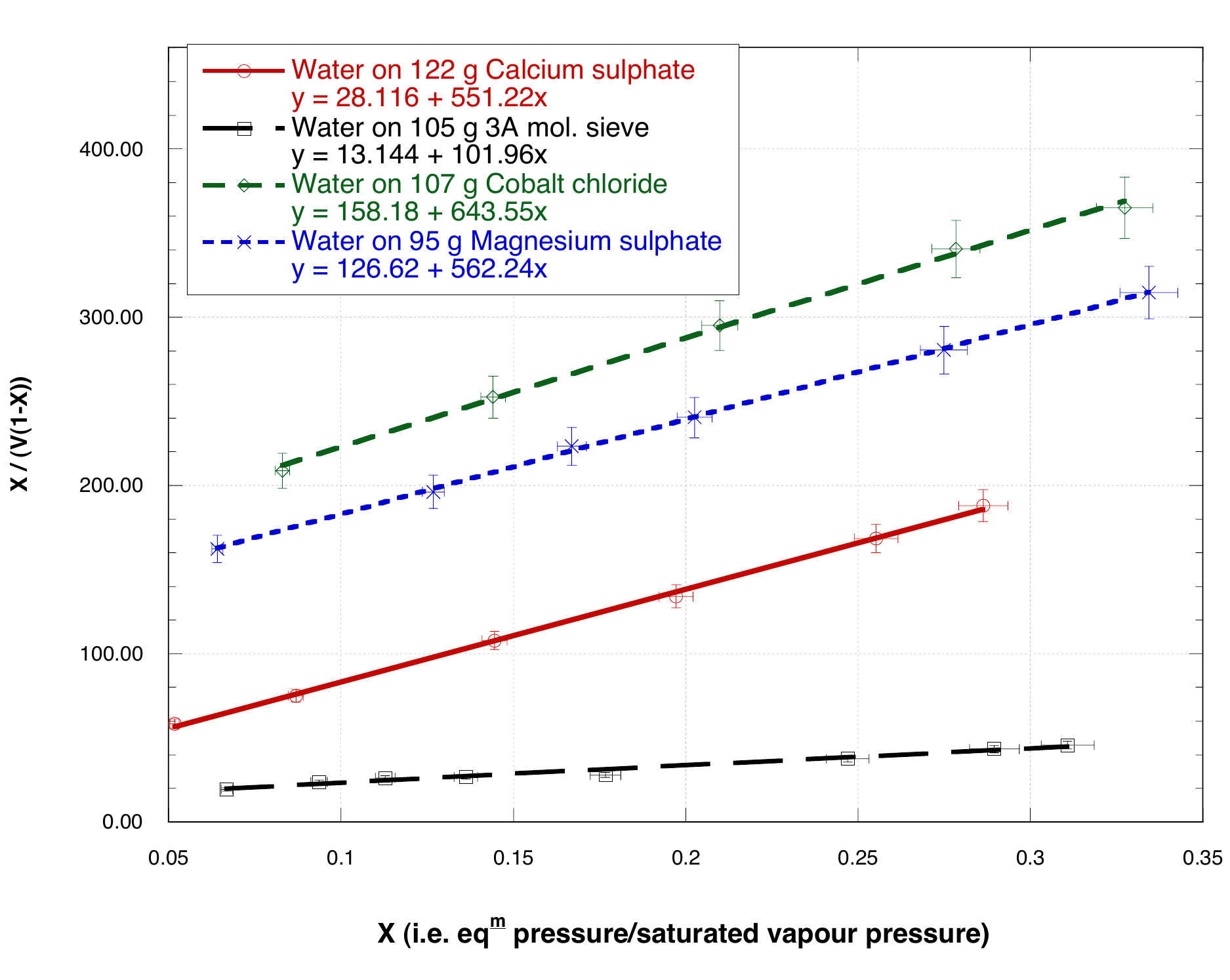}
\end{tabular}
\end{center}

\caption{Adsorption of water vapour on a range of adsorbers and molecular sieves.}
\label{Water_BET2}
\end{figure}

\begin{figure}[t]
\begin{center}
\begin{tabular}{c}
	\includegraphics[width=.8\textwidth]{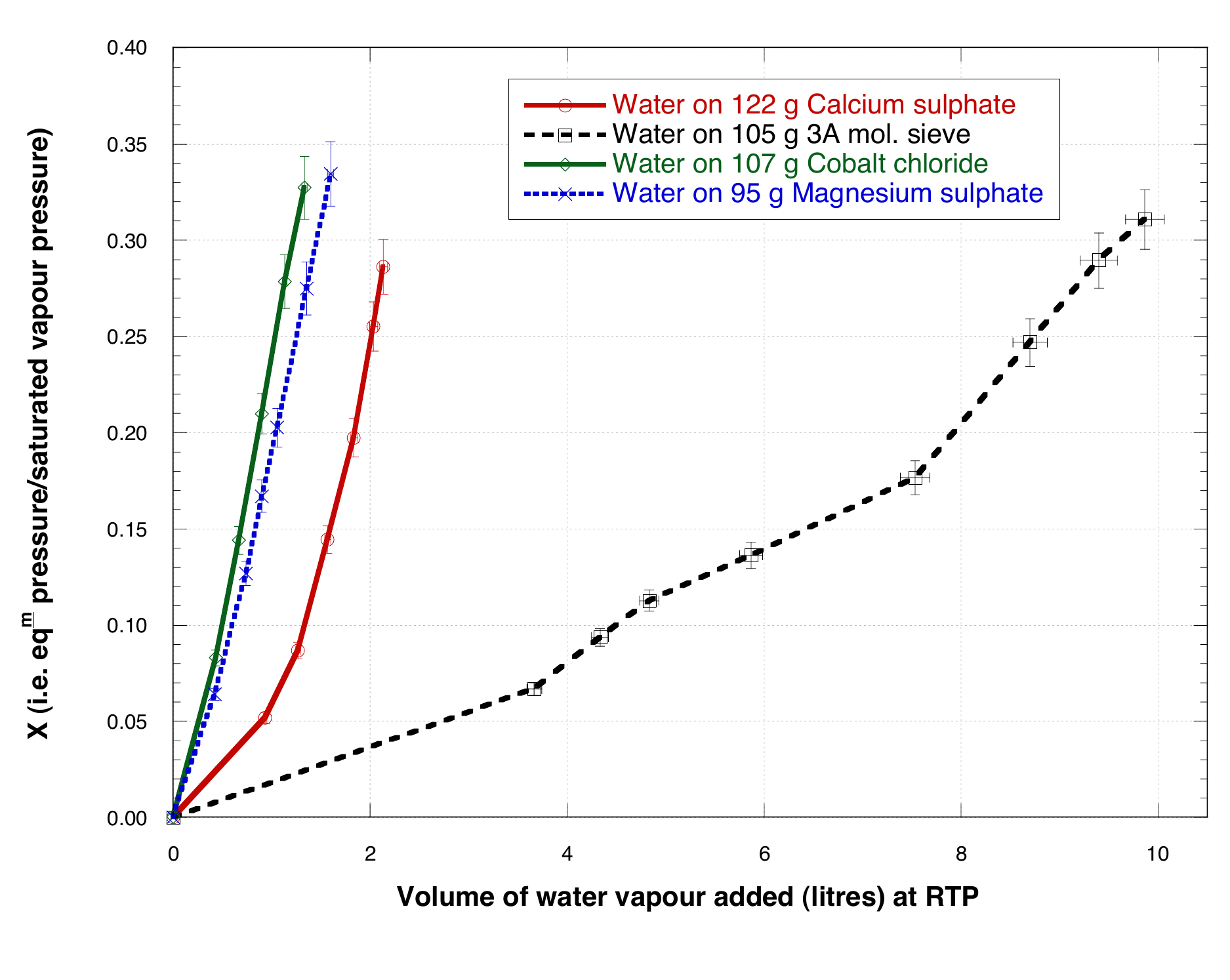}
\end{tabular}
\end{center}

\caption[Adsorption of water vapour on a range of adsorbers and molecular sieves
as a function of added water]
{Adsorption of water vapour on a range of adsorbers and molecular sieves
as a function of added water. Lines are present to illustrate trends and do not indicate
a fit to the data.}

\label{Water_BET1Vol}
\end{figure}

The results of the experiment are shown in Figures~\ref{N2_ArBET1} to~\ref{Water_BET1Vol}.
Linear regression analysis of the data in  Figures~\ref{N2_ArBET1} and~\ref{Water_BET2}  allowed the calculation of
 V$_{m}$ and C values.
Plotting the volume of gas adsorbed as a function of the equilibrium pressure of gas 
above the sample~(Figures~\ref{N2_ArBET1VOL} and~\ref{Water_BET1Vol}) illustrates the rate of gas adsorption.
For each adsorber, the surface area was determined from V$_{m}$ and the molecular area of the adsorbant
as explained above. 
A value for E$_{1}$, the enthalpy of adsorption, and the maximum capacity were also determined and all 
results are shown in Table~\ref{tab:adsorption_results}.

The data 
illustrate that nitrogen and water can be removed very efficiently with the use of
molecular sieves. 
Anhydrous complexes such as calcium sulphate, magnesium sulphate and cobalt chloride,
which do not interact with argon, were found to be much less efficient at removing water compared to
molecular sieves.
Argon is as likely as nitrogen or water to be adsorbed by any
of the molecular sieves when tested individually. 
However, as described in Section~\ref{mol_sieves}, within a mixture of gas where all components can 
enter the pores,
the molecular sieves action is based on thermodynamic selectivity.
Additional substantiation can be found in~\cite{zeolite} where adsorption measurements of air showed a selectivity 
of N$_{2}$ over O$_{2}$ in the same zeolite structure. 
Furthermore, a theoretical prediction in~\cite{Miller1987} reports an adsorption preference 
of N$_{2}$ over O$_{2}$ and Ar in a N$_{2}$, O$_{2}$, Ar gas mixture for a 5A molecular sieve.
Therefore, this selectivity property of molecular sieves among 
components with similar critical diameters allows their use 
for the purification of argon.


\clearpage
%%===========Adsorption results table ==============
\begin{table}
\caption[Adsorption results.] {Adsorption results for the BET isothermal experiment.}

\begin{center}
\vspace{4mm}
{\footnotesize \begin{tabular}{lccc}
\toprule
Combination & \minitab[c]{Enthalpy of\\ adsorption (kJ/mol)} 
& \minitab[c]{Surface\\ area (m$^{2}$)} & \minitab[c] {Maximum\\ capacity (\%wt)} \\
\midrule
N$_{2}$ and activated carbon & 9 & 65800 & 37 \\
Ar and activated carbon & 11 & 62500 & 42\\
N$_{2}$ and 3A Mol. sieve & 11 & 26300 & 23 \\
Ar and 3A Mol. sieve & 11 & 29500& 27 \\
N$_{2}$ and activated alumina & 10 &17300 & 13 \\
Ar and activated alumina  & 11 & 16300 & 16 \\
H$_{2}$O and calcium sulphate & 48 & 4500 & 6 \\
H$_{2}$O and 3A Mol. sieve & 46 & 22800 & 14\\
H$_{2}$O and cobalt chloride & 45 & 3300 & 3 \\
H$_{2}$O and magnesium sulphate & 45 & 3800 & 4 \\
\bottomrule
\end{tabular}}
\end{center}
\label{tab:adsorption_results}
\end{table}

\section{Efficiency of Cu and P$_{2}$O$_{5}$ at Removing O$_{2}$ and H$_{2}$O}
\label{CuPO5}
In this section the efficiency of copper powder and phosphorous pentoxide at the 
removal of oxygen and water respectively from argon gas is evaluated.

Copper reacts strongly with oxygen forming copper oxide. The chemical
reaction can be written as:
\begin{equation}
 \mathrm{2Cu + O_{2} \rightarrow 2CuO} 
\end{equation}

Phosphorous pentoxide with an empirical
formula P$_{2}$O$_{5}$ and molecular formula P$_{4}$O$_{10}$ reacts exothermically 
very strongly with water producing
orthophosphoric acid. The chemical reaction can be written as:
\begin{equation}
\mathrm{P_{4}O_{10} + 6H_{2}O \rightarrow 4H_{3}PO_{4}}
\end{equation}

Both of the chemicals were transferred into the experimental apparatus within an argon
environment to avoid reactions with atmospheric components.
P$_{2}$O$_{5}$ was handled with great care as it is hazardous upon skin contact
and inhalation. In addition, Cu produces a large amount of heat when it reacts with O$_{2}$.
P$_{2}$O$_{5}$ was supplied by Sigma Aldrich~\cite{sigmaaldrich}, whereas
Cu was produced by reducing CuO with H$_2$  at 220 \degr C in a dedicated apparatus.

\subsection{Experimental Method} \label{expmeth_CuPO5}

\begin{figure}[tp]
\begin{center}
\begin{tabular}{c}
	\includegraphics[width=.52\textwidth]{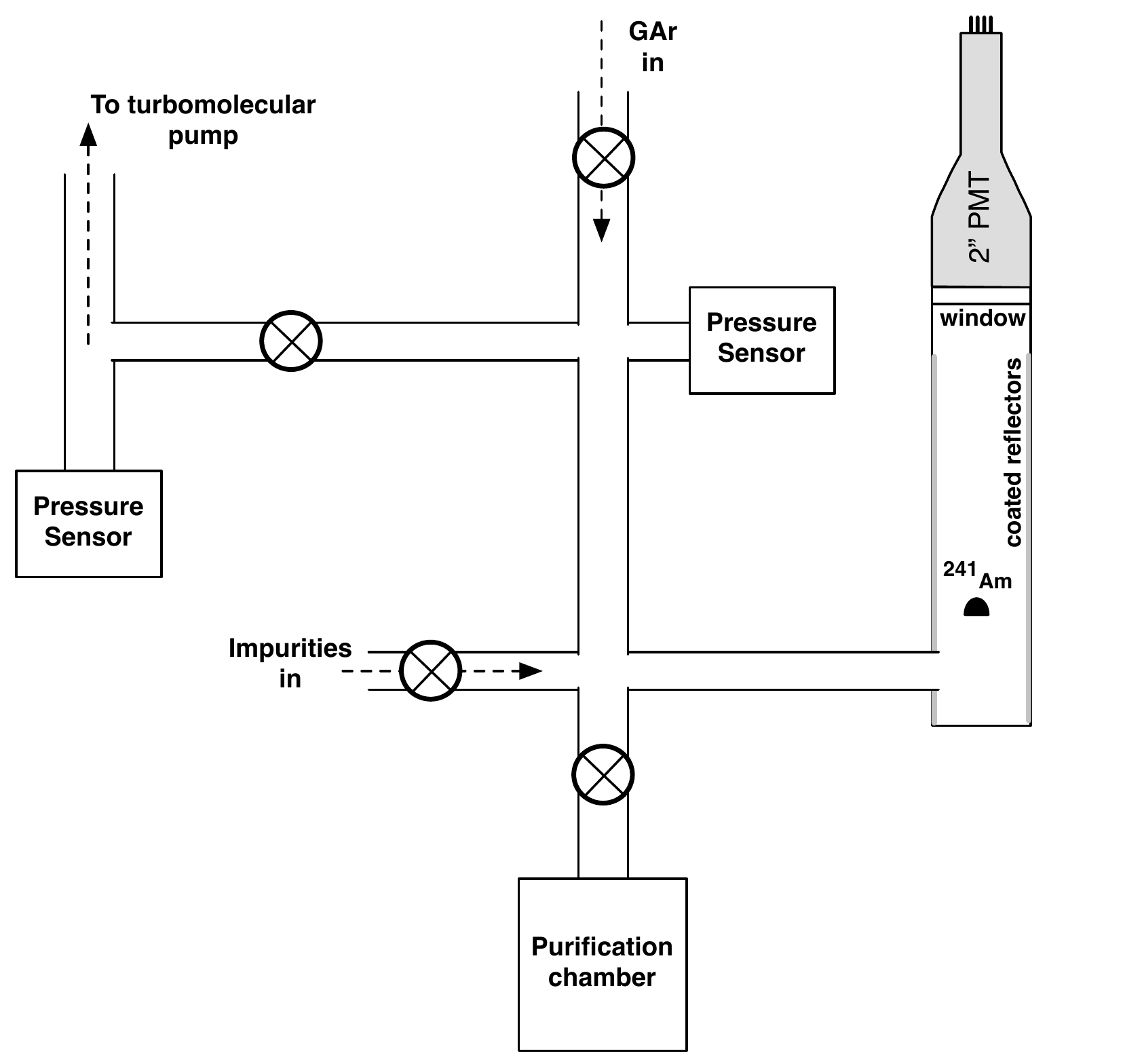}
\end{tabular}
\end{center}

\caption[Schematic of the experimental apparatus for purification tests
using P$_{2}$O$_{5}$ and Cu] 
{A schematic of the experimental apparatus used to asses the efficiency of P$_{2}$O$_{5}$ and Cu
at removing water and oxygen respectively from argon gas. Diagram is not to scale. }

\label{shem:PO5setup}
\end{figure}

A schematic of the experimental apparatus is shown in
Figure~\ref{shem:PO5setup}.
The purifying chemical was placed into a chamber which could be isolated from the rest of the system by a valve.
The system was evacuated to 10$^{-7}$~mbar, the purification chamber was
isolated from the pumping stage and remainder of the apparatus,
and impurities were introduced into the system. The impurities were evacuated to the required partial
pressure (typically 0.5~mbar) and N6 argon gas was added so that the final gas pressure
was 1~bar.
 Purity was then measured using a 200~mm long DN40 tube featuring 1~mg/cm$^{2}$ TPB coated 3M{\small\texttrademark}-foil
 walls with an alpha source at the base and a 2-inch ETL 9831KB PMT fixed externally to a viewing
  window at the top.
 The purification chamber valve was then opened exposing the argon and impurity gas mixture 
 to the purifying chemicals. The increase in purity was then monitored by recording the change of the 
 argon scintillation slow component with time, since  this had been originally calibrated for each
 impurity~(see Figure~\ref{tau2impur}).
 This procedure was repeated at room temperature and -130~$^{\circ}$C for
 Cu and P$_{2}$O$_{5}$ individually and a mixture of Cu and P$_{2}$O$_{5}$ for various partial 
 pressures of oxygen and water.
For the low temperature tests the purification chamber was immersed within
an isobutyl-alcohol/liquid nitrogen mix  
freezing the alcohol to -130~$^{\circ}$C$~\pm$~10~\%. 
Low temperature tests were performed since the chemical reaction rate is also temperature dependent and the purification compounds could potentially be located within the liquid argon target.
 
%\clearpage
\subsection{Results \& Conclusions}

\begin{figure}[b]
\begin{center}
\begin{tabular}{c}
	\includegraphics[width=.7\textwidth]{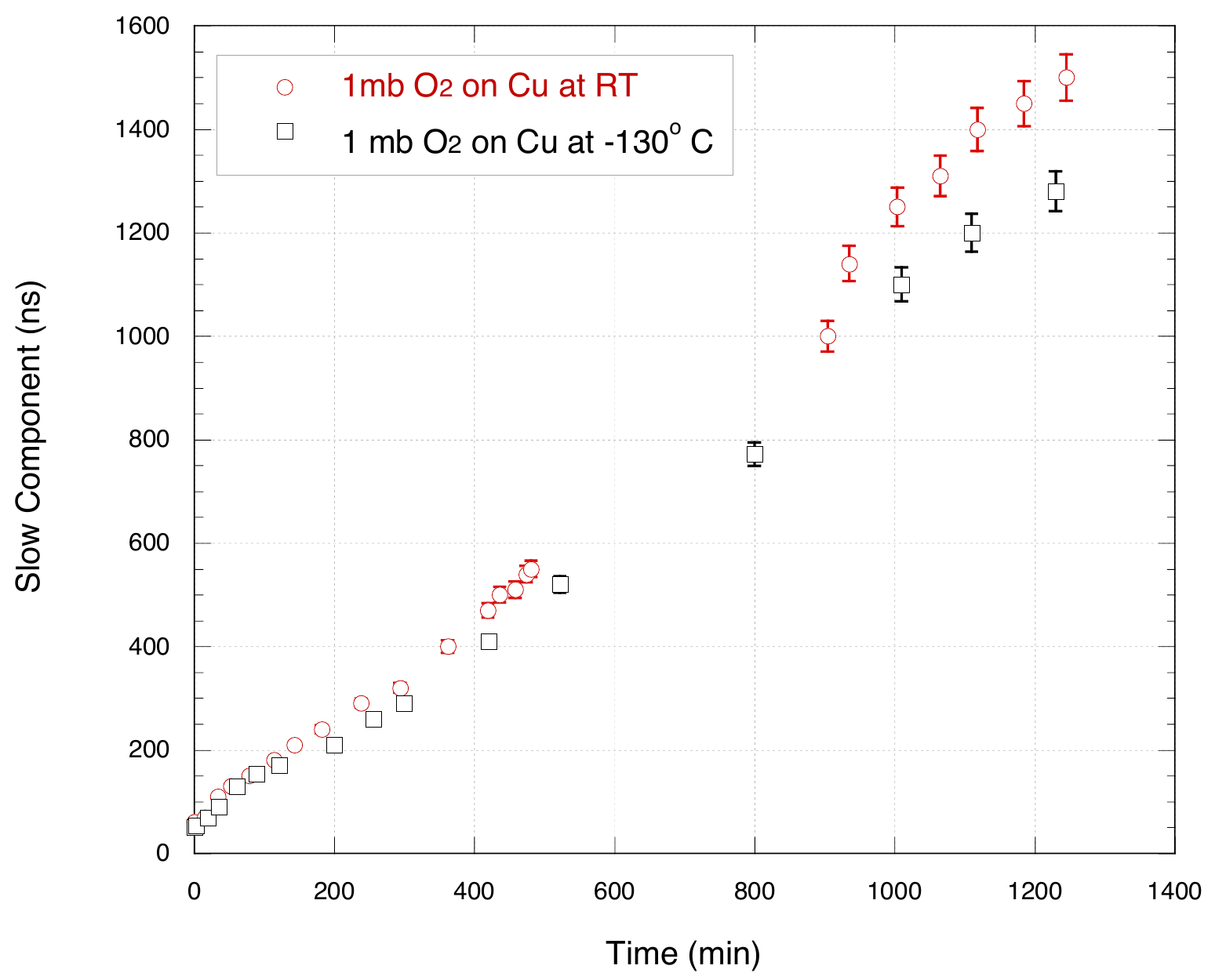}
\end{tabular}
\end{center}

\caption[Cu efficiency at RT and -130~$^{\circ}$C removing O$_{2}$]
{Efficiency of copper at room temperature and -130~$^{\circ}$C at
removing 1~mbar O$_{2}$ partial pressure impurity in 1 bar N6 argon gas.}

\label{1mbO2onCuRTcold}
\end{figure}

Figure~\ref{1mbO2onCuRTcold} shows the slow component decay time
increasing over time when copper was used to remove
1~mb O$_{2}$ partial impurity from 1~bar N6 argon gas
at room temperature and -130~$^{\circ}$C. 
Within approximately 20 hours the slow component decay time 
increased from 60~ns to 1400~ns, indicating that copper removed the O$_{2}$
impurity with high efficiency at both temperatures.

\begin{figure}[tp]
\begin{center}
\begin{tabular}{c}
	\includegraphics[width=.7\textwidth]{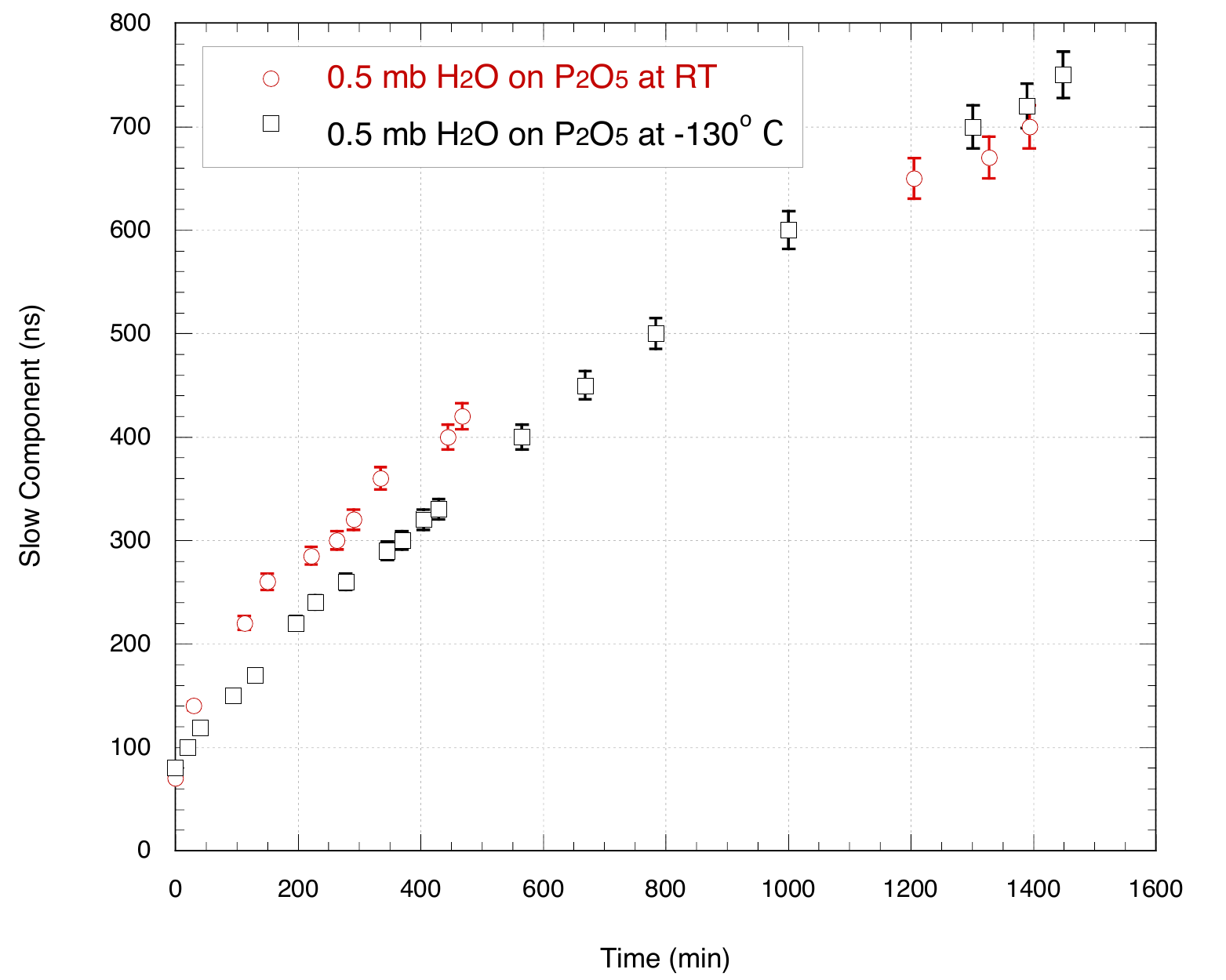}
\end{tabular}
\end{center}

\caption[P$_{2}$O$_{5}$ efficiency at RT and at -130~$^{\circ}$C
removing H$_{2}$O]
{Efficiency of phosphorous pentoxide at room temperature and -130~$^{\circ}$C
at removing 0.5~mbar H$_{2}$O partial pressure impurity in 1 bar N6 argon gas.}

\label{05mbH2O_PO5_RT_cold}
\end{figure}

Phosphorous pentoxide also very capably improved the argon gas purity (down to a level of $\sim$ 4~ppm O$_{2}$
equivalent impurity, based on Figure~\ref{tau2impur})
by removing
0.5~mb H$_{2}$O at room temperature and -130~$^{\circ}$C as shown in
Figure~\ref{05mbH2O_PO5_RT_cold}.

\begin{figure}[tp]
\begin{center}
\begin{tabular}{c}
	\includegraphics[width=.7\textwidth]{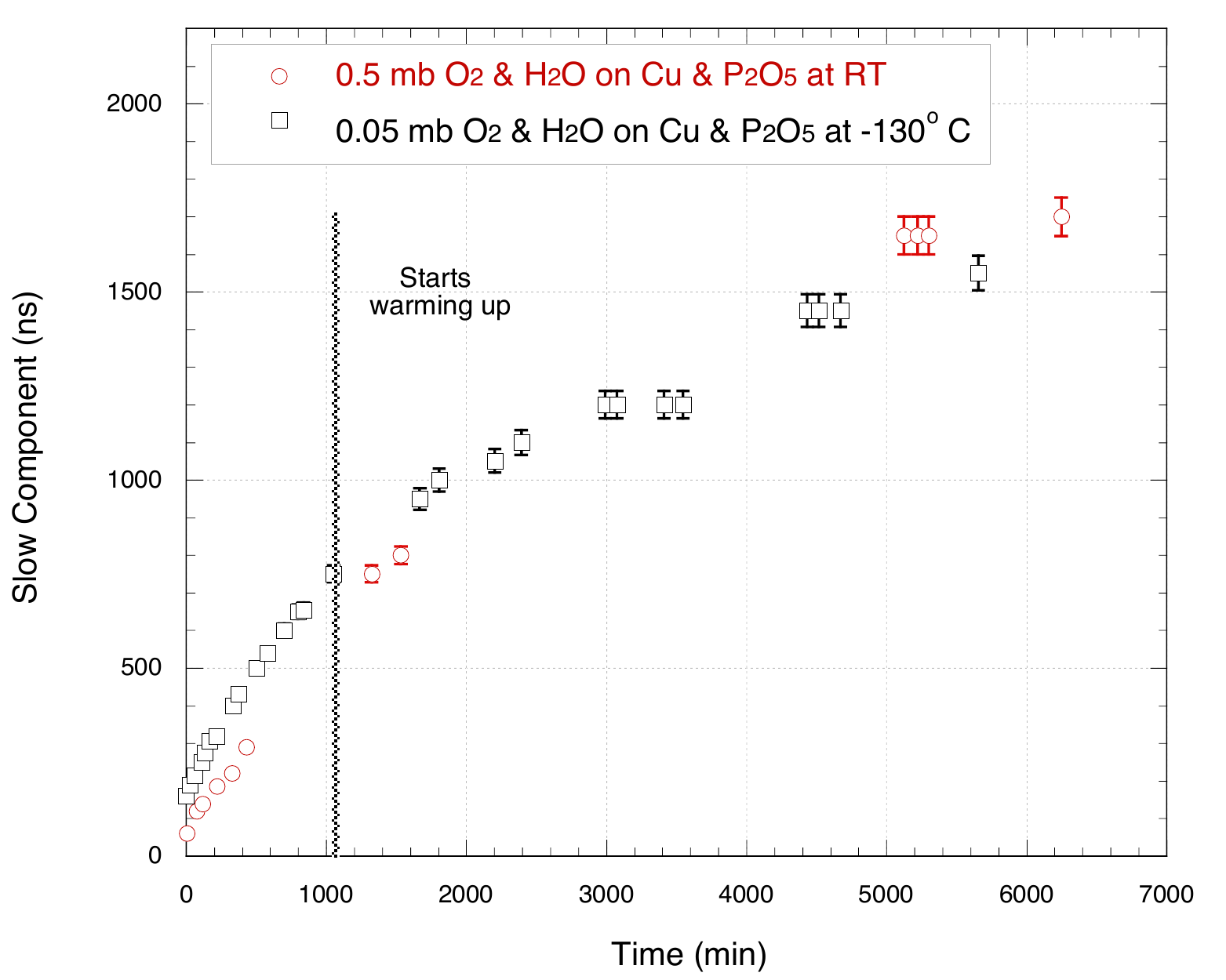}
\end{tabular}
\end{center}

\caption[P$_{2}$O$_{5}$ and Cu mixture efficiency at RT and -130~$^{\circ}$C
at removing H$_{2}$O and O$_{2}$] 
{Efficiency of copper and phosphorous pentoxide mixture at both room temperature and -130~$^{\circ}$C
at removing 0.5~mbar and 0.05~mbar respectively O$_{2}$/H$_{2}$O partial pressure impurity in 1 bar
N6 argon gas.
The vertical black line indicates the point at which the purification mixture started to warm up.}

\label{05_005mbO2H2O_CuPO5_RT_cold}
\end{figure}

In addition, rapid purity improvement was observed for the Cu, P$_{2}$O$_{5}$ mixture 
during the first 17~hours of experimental operation
at both temperatures tested (see Figure~\ref{05_005mbO2H2O_CuPO5_RT_cold}).   
The vertical black line in Figure~\ref{05_005mbO2H2O_CuPO5_RT_cold}
denotes the point at which the LN$_{2}$ supply was stopped and 
the purification mixture was left overnight while gradually reaching room temperature~(it is estimated
that room temperature was reached within 5 hours).
After 30 hours of experimental operation the slow component increased from approximately 50~ns
to 1000~ns with this mixture and after 70 more hours the slow component had reached 1650~ns.
With reference to Figure~\ref{tau2impur}, 1650~ns corresponds to approximately 3~ppm
O$_2$ equivalent impurity. 1650~ns purity appears to be the purification limit of this mixture. 
This could be explained by the fact that Cu and  P$_{2}$O$_{5}$ only remove O$_2$ and H$_{2}$O impurities,
thus small concentrations of N$_{2}$ and CO$_{2}$ introduced with the O$_2$ (the O$_2$ used 
had 0.5~\% impurities) remain in the system.

When comparing room and low temperature data for all the mixture combinations tested, it was 
observed that there was no significant alteration in the rate of purity increase, although a reduction in the rate of reaction was expected at the lower temperature.
In the case of purification involving water impurities
this may be due to the fact that at low temperatures the water
condensed on the walls of the apparatus and froze,
thus allowing a faster increase of purity while the actual reaction rate
was slower. 
In the case of the copper reaction with O$_{2}$,
this could be attributed to the longer residency time of O$_{2}$ on the copper's surface at low temperatures.
In other words, the molecules of O$_{2}$ move slower at low temperatures, thus staying
for a longer period on the surface of the copper powder giving it more time to react.

%%%--------------------

\clearpage

%%%%---------------------
\clearpage

\section{The Liverpool LAr Setup }
\label{LiverpoolSetup}

\begin{figure}[ht]
\begin{center}
\begin{tabular}{c}
	\includegraphics[width=.6\textwidth]{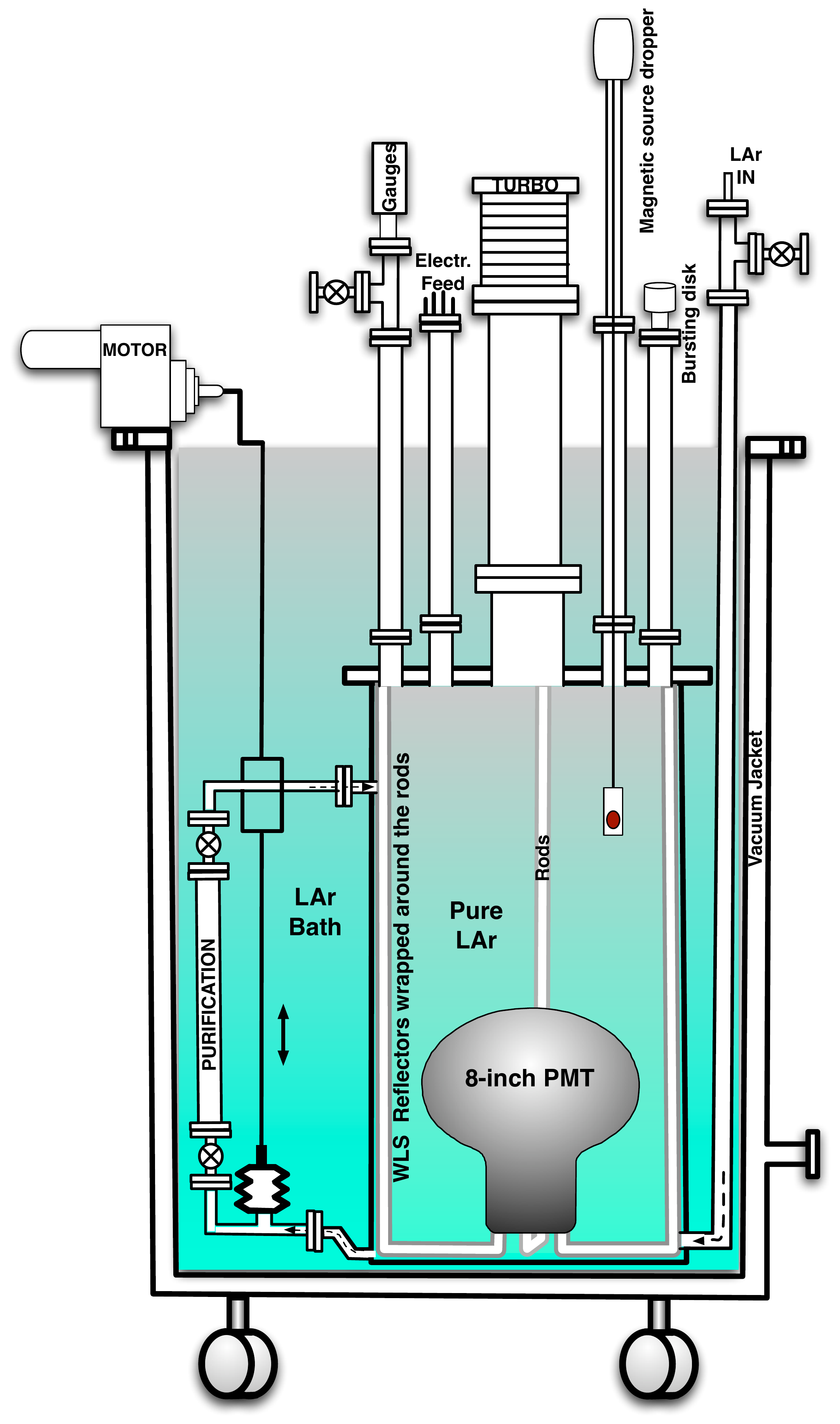}
\end{tabular}
\end{center}
\caption{A schematic of the Liverpool LAr setup.}

\label{fig:LiverpoolSetup}
\end{figure}

The Liverpool LAr setup, shown in Figure~\ref{fig:LiverpoolSetup}, consists of a 40 litre
 stainless steel conflat vacuum target vessel, 
which contains all the detector components and has an externally connected purification system on one side. 
The top flange has 
five vacuum ports used to mount a turbomolecular pump, pressure gauges, electrical feedthroughs, a bursting disk, and
a movable magnetic actuator. 
The 40 litre vessel sits  within a 250 litre stainless steel open LAr bath,
in order to maintain the cryogenic argon temperature. The bath vessel, constructed by the University of Liverpool mechanical workshop, consists of  two cylindrical chambers placed 2~cm apart concentrically and argon welded to the top flange, thus creating a vacuum jacket in order to minimise heat losses by conduction and convection. The outside of the inner bath chamber was wrapped twice with Mylar reflector to minimise heat losses by radiation. 

The internal detector assembly consists of an 8-inch Pt underlay cryogenic Hamamatsu R5912-02MOD PMT, 
held at the bottom facing upwards by three stainless steel rods, around which
3M{\small\texttrademark}-foil reflectors are wrapped.
To shift the argon VUV light to the high quantum efficiency region of the PMT,
the PMT window and 3M{\small\texttrademark}-foil reflectors were coated with 0.05~mg/cm$^2$ 
and 1~mg/cm$^2$ TPB thickness respectively via vacuum evaporation.
A parallel plate capacitor, positioned at the side of the PMT, was used
as a LAr level sensor.
The signal of the PMT was digitised at a sampling rate of 1 GS/s, using an Acqiris DP1400 digitiser.
High rate scintillation light was generated with an Am-241 alpha source positioned on the magnetic actuator, 
which allowed 
the source to be moved between the liquid and gas phase of the detector.
%%---------
%%-------
%%---------

\subsection{The Purification System}

\begin{figure}[ht]
\begin{center}
\begin{tabular}{c}
	\includegraphics[width=.7\textwidth]{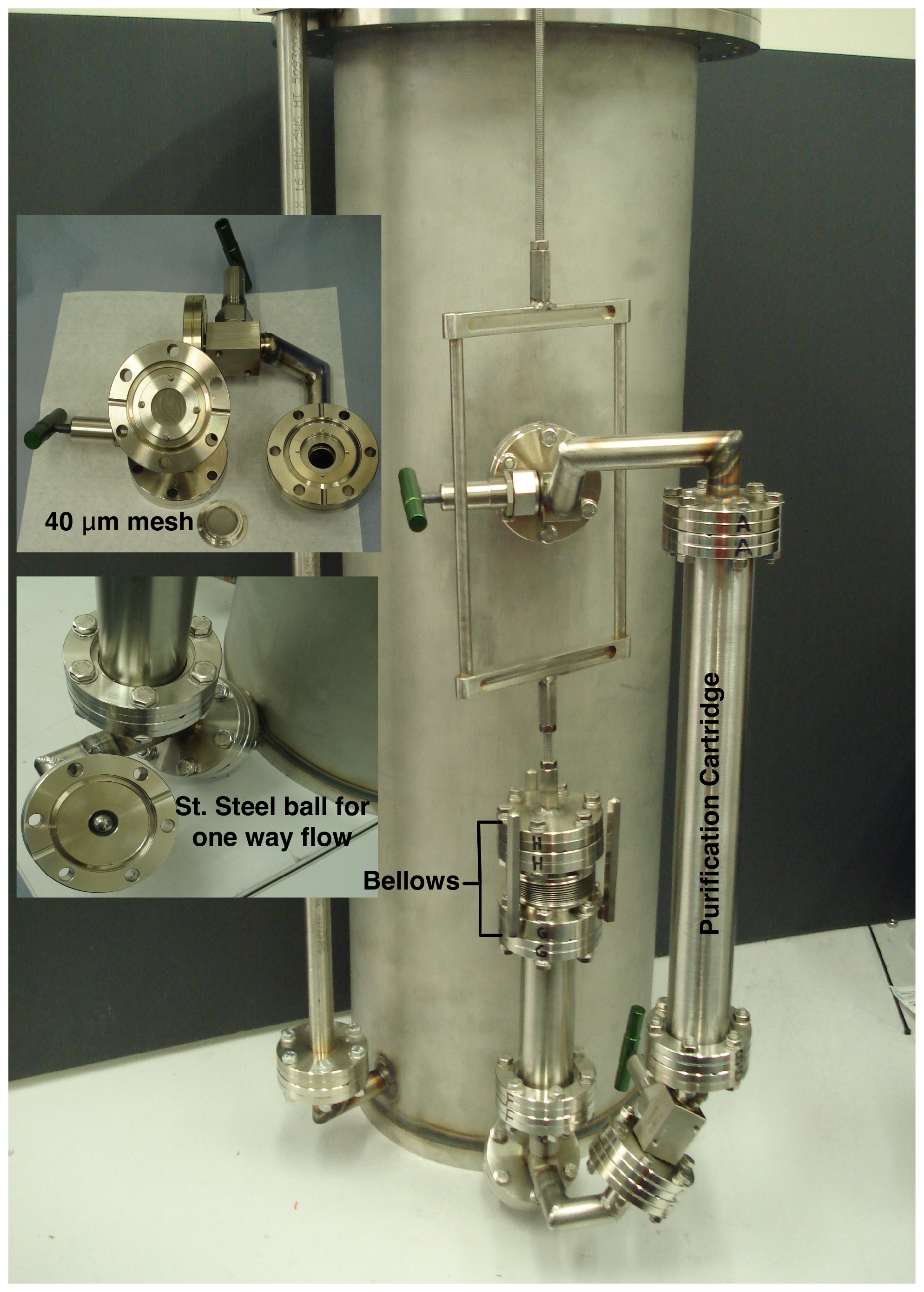}
\end{tabular}
\end{center}
\caption{The target vessel and re-circulation system.}

\label{fig:recirculation}
\end{figure}

The purification system is mounted on the side of the 40~litre vessel. LAr re-circulation from the bottom of the target vessel and through a purification cartridge is achieved using an all-metal motorised bellows pump.
This bellows pump (Figure~\ref{fig:recirculation}), manufactured in Liverpool
using high yield low modulus steel, operates with effectively zero 
wear, zero friction internal components, producing a sustained throughput across a pressure gradient 
during expansion and compression. 
A one way valve system was achieved by placing stainless steel balls before and after the bellows
junction.
The bellows is operated by an external geared motor to which it is connected
via a connecting rod.
As the bellows is completely immersed in the LAr bath, the only heat load is 
along the connecting rod to the external geared motor.

The re-generatable purification cartridge contained 250~g of activated copper and 300~g of a 3A, 4A and 13X molecular sieve mixture, all purchased from Sigma Aldrich~\cite{sigmaaldrich}.
The chemicals were contained within the cartridge by a 40 $\mum$ stainless steel mesh
at either end. 
In this experiment phosphorous pentoxide was omitted from the cartridge due to safety
concerns, and was replaced by the molecular sieves which were found to 
be efficient at adsorbing water.
  
The copper can be re-generated
by the application of 220~\degr C using a heating tape wrapped around the cartridge 
and flushing with a mixture of 70~\% argon and 30~\% hydrogen gas. Once the copper is activated,
re-baking the cartridge at 250~$^o$C under vacuum for $\sim$8~hours 
will result in the reactivation of the molecular sieves.

%%-------
%%---------
\subsection{Demonstration of Re-circulation and Purification}
\begin{figure}[ht]
\begin{center}
\begin{tabular}{c}
	\includegraphics[width=.85\textwidth]{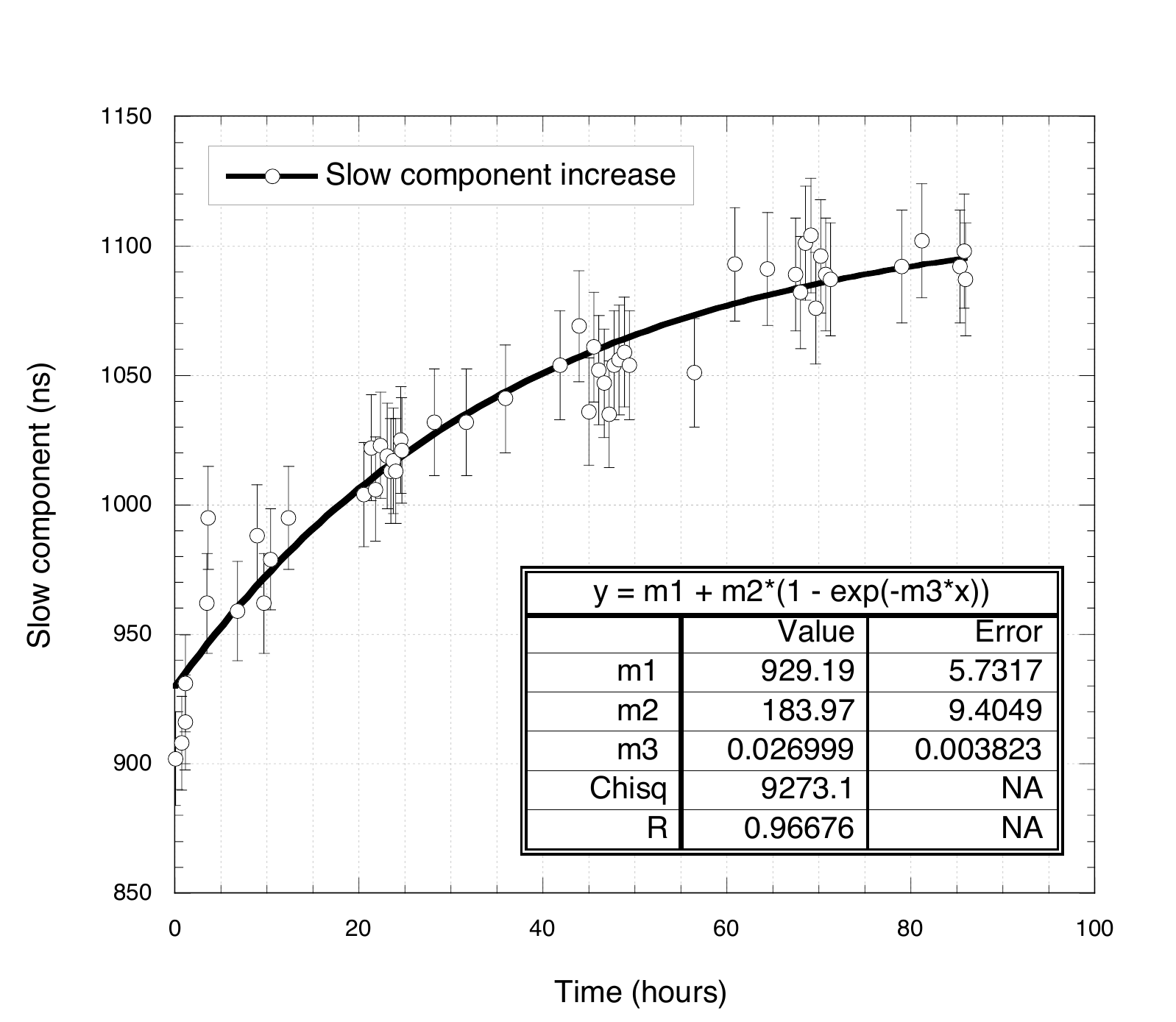}
\end{tabular}
\end{center}
\caption{Purity increase over four days continuous re-circulation through the cartridge at rate of 27~litres/hour.}

\label{fig:Purityrecirc}
\end{figure}

The gear motor was powered by a variac transformer adjusted to 30~rpm. 
The pump was successfully operated in liquid argon for 4 days at a pump rate of 30~rpm~(0.5~Hz) without
 any failure or indication of degradation due to metal fatigue. 
At a pump rate of 30~rpm a single cycle displaces 15~ml, thus giving a flow rate of 27~litres/hour.
Therefore, the entire LAr volume was re-circulated in less than two hours.
The scintillation slow component
decay time was used to monitor the purity during re-circulation. Figure~\ref{fig:Purityrecirc}
 shows the slow component increase from approximately 900~ns to 1100~ns
 over a period of four days of re-circulation.

%%---------

\subsection{Effect of O$_{2}$ Impurity on LAr Scintillation}
\label{contamLAr}
\begin{figure}[ht]
\begin{center}
\begin{tabular}{c}
	\includegraphics[width=.5\textwidth]{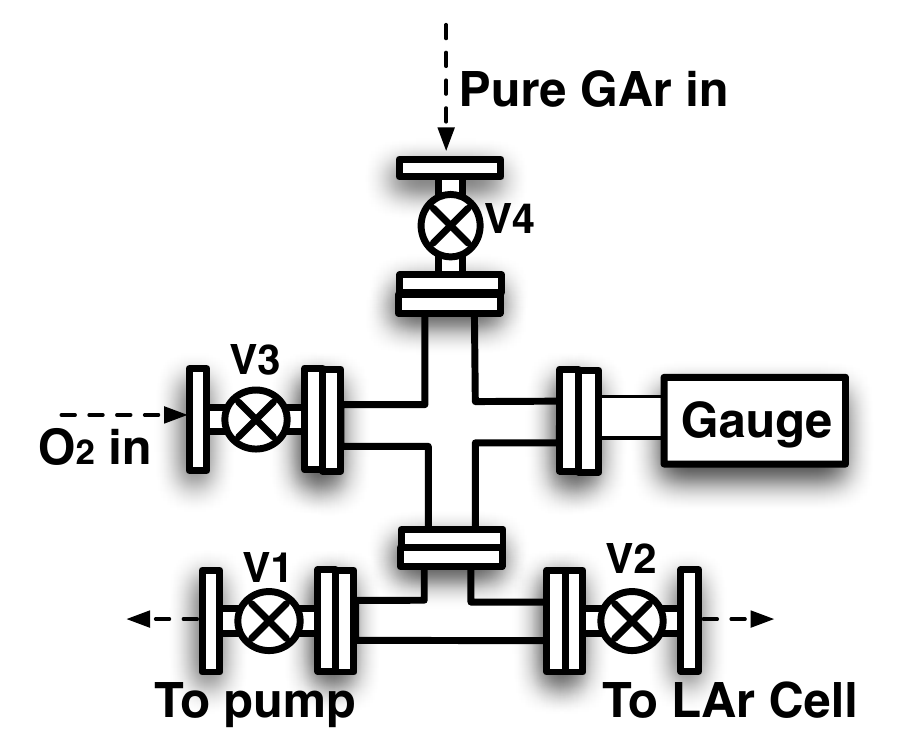}
\end{tabular}
\end{center}
\caption{Schematic of the apparatus used to introduce specific amounts of O$_2$ impurity.}

\label{fig:contamination_mixer}
\end{figure}

An additional experiment using the Liverpool LAr setup was performed in order to investigate the effect on
the LAr scintillation light of 0.01~ppm to 100~ppm O$_2$ contamination levels.
Known partial pressures of O$_2$ impurity were introduced into the system
by employing a dedicated apparatus as shown in Figure~\ref{fig:contamination_mixer}.
As the concentration of O$_2$ was increased, PMT data were recorded and analysed for the total pulse area, and slow component decay time.

Prior to filling, the target vessel~(including the contamination apparatus) was evacuated to 10$^{-7}$~mb using a turbomolecular pump.
1 bar of pure argon gas was added into the system which was then cooled down at a rate of 1~\degr C/min by the slow addition of  LAr into the bath, thus avoiding any thermal shock on the PMT. Once the target vessel reached LAr temperature, it was filled with 20~litres of LAr by condensing argon gas withdrawn
from the gas phase of a LAr dewar purchased from BOC~\cite{BOC}.
A first level of purification was made by passing the GAr through a purification getter containing 300~g of 3A and 13X molecular sieves before
entry into the vessel. 
The initial slow component  of the GAr introduced was measured to be approximately 3200~ns indicating a gas 
purity of $<$1~ppb based on the measurements described in Section~\ref{impurities_effect}.
Once the target vessel was full, purification in the liquid phase was initiated by operation of the all-metal bellows pump recirculation system. After approximately 24~hours re-circulation, the LAr slow component decay time increased by 100~ns and levelled off at approximately 1100~ns based on Am-241 data.

 Contamination of the target vessel with the required partial pressures, equivalent to 0.01~ppm, 0.1~ppm, 0.2~ppm,
 0.4~ppm, 0.8~ppm, 1~ppm, 2~ppm, 4~ppm, 8~ppm, 10~ppm, 20~ppm and 100~ppm, were introduced  using the
 dedicated apparatus (Figure~\ref{fig:contamination_mixer}) of 0.52~litres  volume.
Each contamination level was introduced by performing the following steps.
The contamination apparatus was isolated from the 
system by closing valve V2, evacuated to 10$^{-3}$~mb, and then filled 
with 1 bar O$_2$ gas\footnote{This step,  involving evacuating the contamination apparatus 
and then filling it with O$_2$, was repeated many times to ensure that when the apparatus was finally
pumped down to 10$^{-3}$~mb, the gas within was almost totally O$_2$ with no unknown impurities.}. 
The apparatus was then pumped down to the required partial pressure, which was then mixed with 1 bar of pure 
argon gas and inserted into the system by opening valve V2. 
Finally, pure argon gas was flushed into the system for about 1 min 
in order to speed up the dilution process of the impurity with the LAr volume. 
As the melting and boiling points of O$_2$ are - 218~\degr C and -183~\degr C respectively,
O$_2$ will have a liquid form in LAr.
To verify uniform dilution, data-sets were
recorded over a one hour time span between consecutive contamination additions.
On average the impurity was observed to be fully dispersed within 20 min.

\begin{figure}[t]
\begin{center}
\begin{tabular}{c}
	\includegraphics[width=.66\textwidth]{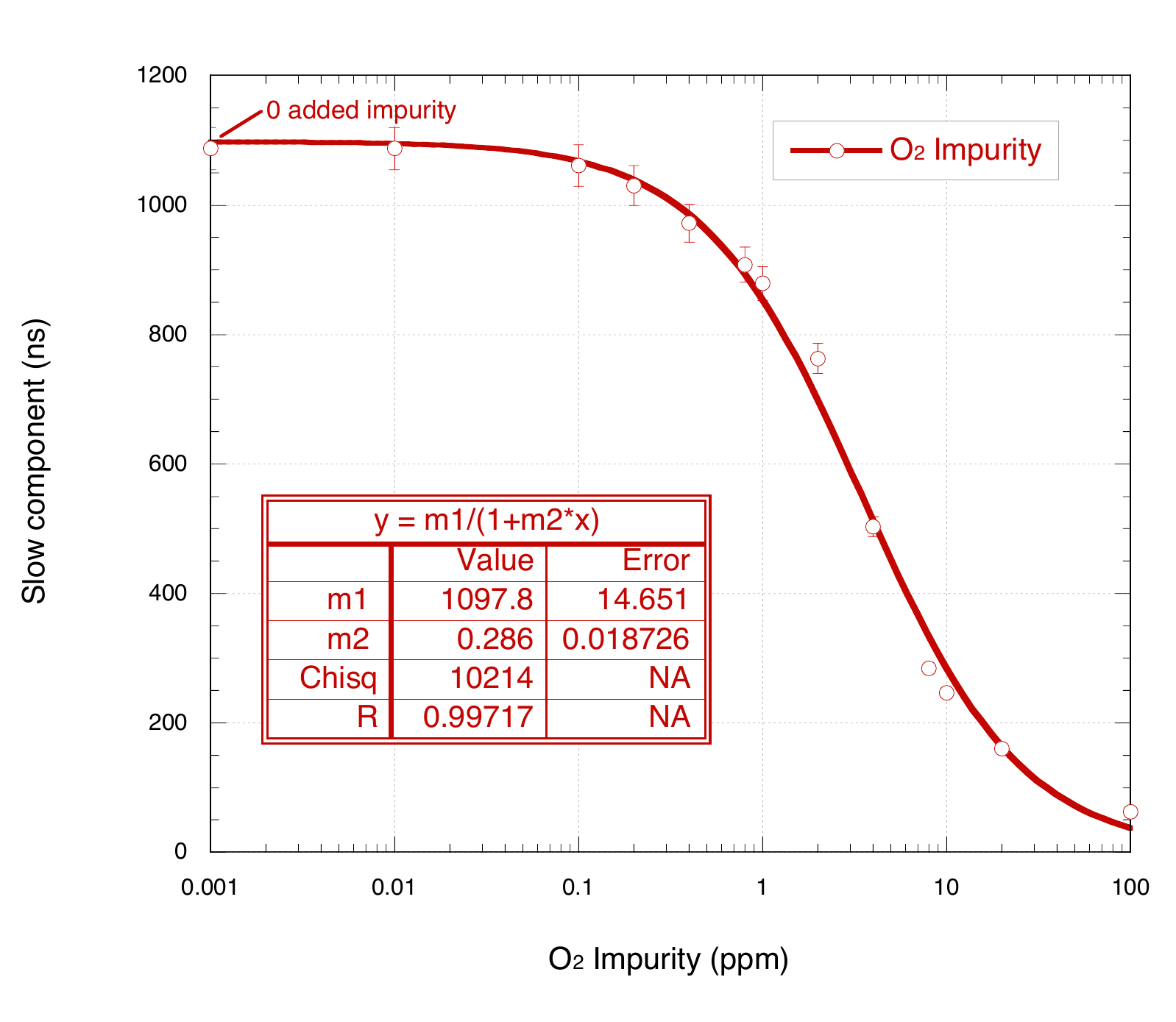}
\end{tabular}
\end{center}

\caption[Slow component (triplet) decay time variation with increasing ppm O$_2$ impurity concentrations]
{Slow component (triplet) decay time variation with increasing ppm O$_2$ impurity concentrations in liquid argon.
The data were fitted with a Birk's law type function, where m1~=~1098~ns$\pm$15~ns 
and m2~=~0.29$\pm$0.02. The slow component observed at 1~ppm impurity corresponds 
well to that observed for liquified N6 argon gas~(Figure~\ref{fig:Purityrecirc}).}

\label{LAr_tau2impur_contam}
\end{figure}

\begin{figure}[t]
\begin{center}
\begin{tabular}{c}
	\includegraphics[width=.66\textwidth]{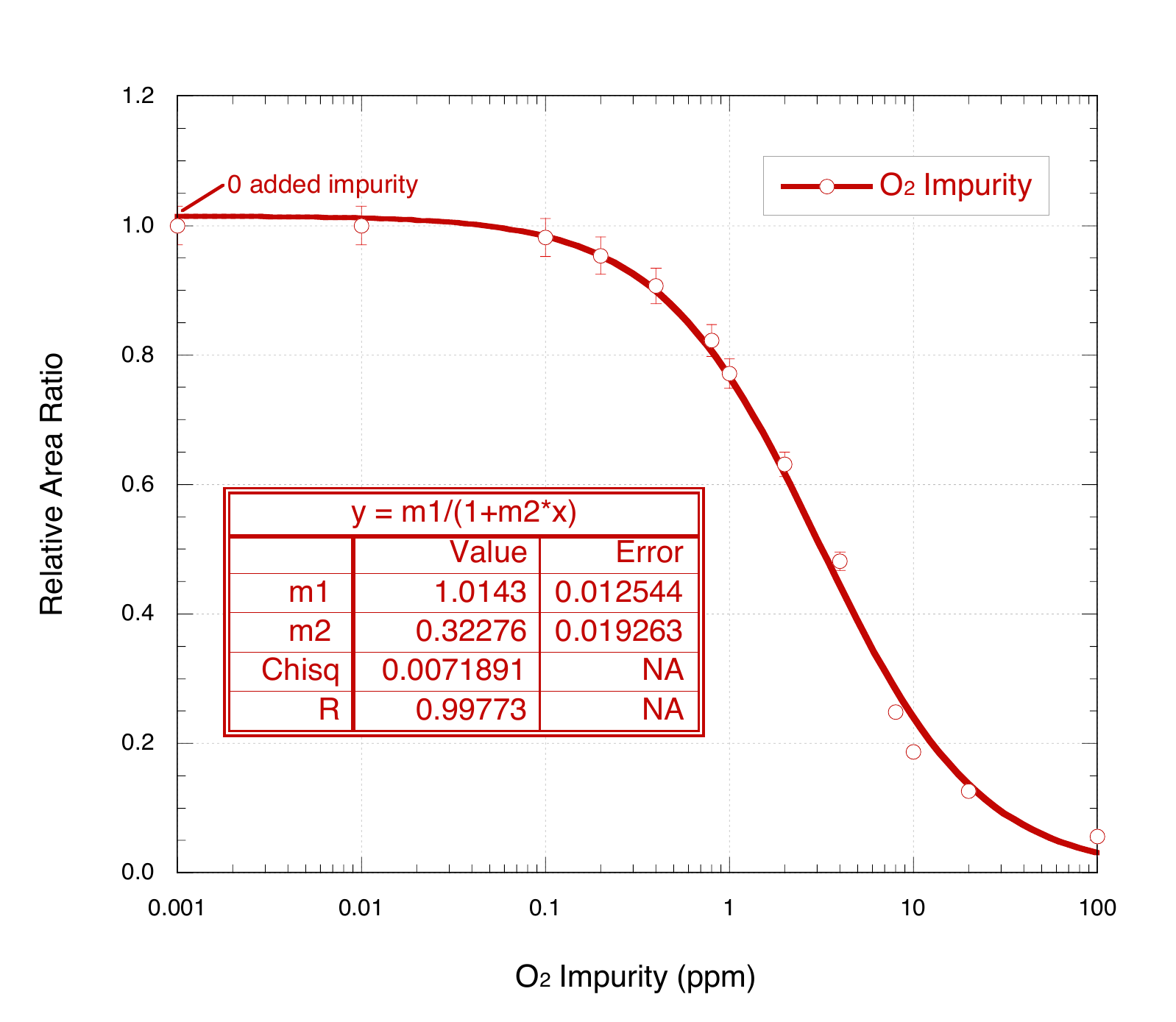}
\end{tabular}
\end{center}

\caption{Ratio of contaminated argon pulse area to pure argon pulse area for increasing ppm O$_2$ impurity concentrations in liquid argon. The data were fitted with a Birk's law type function, where m1~=~1.01$\pm$0.01 
and m2~=~0.32$\pm$0.02.}

\label{LAr_QF_contam}
\end{figure}

For all impurity levels, 20000 events with a 10 $\mu$s time window were recorded.
 Each individual event pulse was parameterised by the sum of two exponential functions
 with all parameters varying in the fit.
 The Gaussian mean of the pulse parameters was then monitored. Figure~\ref{LAr_tau2impur_contam} shows the correlation of the slow component decay time with
ppm added O$_2$ impurity, whereas Figure~\ref{LAr_QF_contam} shows the ratio of the contaminated total pulse area to pure argon total pulse area. 
A Birks' law type function fit (similarly to Section~\ref{impurities_effect} for gas data)  was applied to the data allowing an approximate description of
the quenching effect of O$_2$
impurity on the slow component and total deposited energy of the LAr scintillation light.

%%---------

%%------------

%%------------

%-------------------------------------------------------------------------
\clearpage
\section{Conclusions}

The study of the effect of a range of impurity partial pressures within
1~bar N6 argon gas on the scintillation light 
showed that the slow component is very sensitive to the impurities tested, with H$_{2}$O
having the greatest effect followed by CO$_{2}$, O$_{2}$ and lastly N$_{2}$.
This allows the decay time of the slow component
to be calibrated to the partial pressure of the impurities utilised.

Adsorption of Ar gas, N$_{2}$ gas and H$_{2}$O vapour by
various molecular sieves and anhydrous complexes was studied.
The molecular sieves were far superior than the anhydrous complexes
at adsorbing the impurities.
The use of molecular sieves for argon purification
will not be compromised by the fact that molecular sieves adsorb argon,
as within a mixture of gas, where all components can enter the pores, 
the action is based on thermodynamic selectivity.

The efficiency of Cu and P$_{2}$O$_{5}$ at removing oxygen and water
impurities from 1~bar N6 argon gas at both room temperature 
and -130~$^{\circ}$C was found to be high.

A novel LAr re-circulation system with a home made purification cartridge and a motorised metal bellows pump
was successfully operated.
Within 4 days of continuous re-circulation  the slow component decay time was increased by 200~ns 
reaching $\sim$1100~ns.
The amount of LAr passing through the purification cartridge can be effectively increased by
using bigger bellows and connecting multiple ones in parallel.
For  example, a 4 cylinder pump with DN63CF bellows could re-circulate over 450 
litres per hour, thus making it applicable and cost effective for a multi tonne scale system.

Finally, the slow component decay time and the total deposited energy of the LAr scintillation light
variation with increasing ppm O$_2$ impurity concentrations (0.01~ppm to 100~ppm) was determined.

\acknowledgments

Part of this work was submitted by K. Mavrokoridis in partial fulfilment for the degree of Doctor of Philosophy, University of Sheffield, UK. 
The Authors are grateful for the expertise and dedicated contributions of the Mechanical Workshop of the Physics
Department, University of Liverpool. We acknowledge the support of the University of Liverpool, STFC, and the Royal Society.

%% ==============================================

%% ==============================================

\end{document}